\documentclass[a4paper,11pt]{article}
\pdfoutput=1 \usepackage{amssymb}
\usepackage{amsmath}
\usepackage{bbold}
\usepackage{bm}
\usepackage{braket}
\usepackage{mathtools}
\usepackage[usenames,dvipsnames,svgnames,table]{xcolor}
\usepackage [utf8] {inputenc}
\usepackage{color}
\usepackage{cite}
\usepackage{subfigure}
\usepackage{lmodern}
\usepackage{footnote}
\usepackage{glossaries-extra}
\setabbreviationstyle[acronym]{long-short}
\glssetcategoryattribute{acronym}{nohyperfirst}{true}

\usepackage{slashed}
\usepackage[pdftex] {graphicx}
\usepackage{multirow}
\usepackage{jheppub}
\usepackage{here}
\usepackage{hyperref}
\usepackage{verbatim}
\usepackage[justification=centering, singlelinecheck=false]{caption}

\newcommand{\cA}{\mathcal{A}}
\newcommand{\cB}{\mathcal{B}}
\newcommand{\cH}{\mathcal{H}}

\newcommand{\cD}[0]{\mathcal D}
\newcommand{\cK}[0]{\mathcal K}
\newcommand{\cL}[0]{\mathcal L}
\newcommand{\cM}[0]{\mathcal M}
\newcommand{\cO}[0]{\mathcal O}
\newcommand{\cR}[0]{\mathcal R}
\newcommand{\cS}[0]{\mathcal S}
\newcommand{\cT}[0]{\mathcal T}

\newcommand{\df}[0]{\mathrm{df}}

\newcommand{\Kdf}[0]{{\cK_{\df,3}}}

\newcommand{\PV}[0]{{\mathrm{PV}}}

\newcommand{\KSS}[0]{Kim:2005gf}

\newcommand{\BHSnum}[0]{Briceno:2018mlh}
\newcommand{\BHSK}[0]{Briceno:2018aml}
\newcommand{\HSQCa}[0]{Hansen:2014eka}
\newcommand{\HSQCb}[0]{Hansen:2015zga}
\newcommand{\dwave}[0]{Blanton:2019igq}
\newcommand{\largera}[0]{Romero-Lopez:2019qrt}

\newcommand{\isospin}[0]{Hansen:2020zhy}
\newcommand{\BSequiv}[0]{Blanton:2020jnm}
\newcommand{\BSQC}[0]{Blanton:2020gha}

\newcommand{\LL}[0]{Lellouch:2000pv}

\newcommand{\BHW}[0]{Briceno:2014uqa}
\newcommand{\BHLL}[0]{Briceno:2015csa}

\newcommand{\Hadspecpipipi}[0]{Hansen:2020otl}
\newcommand{\Jackurainteq}[0]{Jackura:2020bsk}
\newcommand{\KT}[0]{Khuri:1960zz}

\newcommand{\ThreeQCDNumerics}[0]{%
Mai:2018djl,
Horz:2019rrn,
Blanton:2019vdk,
Culver:2019vvu,
Mai:2019fba,
Fischer:2020jzp,
Hansen:2020otl,
Alexandru:2020xqf,
Brett:2021wyd}

\newcommand{\HVPfinitevolume}[0]{%
Bernecker:2011gh,
Meyer:2018til}

\newcommand{\ThreeBody}[0]{%
Briceno:2012rv,
Polejaeva:2012ut,
Hansen:2014eka,
Hansen:2015zga,
Briceno:2017tce,
Hammer:2017uqm,
Hammer:2017kms,
Mai:2017bge,
Briceno:2018aml,
Briceno:2018mlh,
Jackura:2019bmu,
Blanton:2019igq,
Briceno:2019muc,
Hansen:2019nir,
Romero-Lopez:2019qrt,
Blanton:2020gha,
Blanton:2020jnm,
Hansen:2020zhy,
Blanton:2020gmf,
Muller:2020vtt}

\newcommand{\transitions}[0]{%
Lin:2001ek,
Detmold:2004qn,
\KSS,
Christ:2005gi,
Meyer:2011um,
Hansen:2012tf,
Briceno:2012yi,
Bernard:2012bi,
Agadjanov:2014kha,
Briceno:2014uqa,
Feng:2014gba,
Briceno:2015csa,
Briceno:2015tza,
Baroni:2018iau,
Briceno:2019nns,
Briceno:2020xxs,
Feng:2020nqj}

\newcommand{\phaseRefs}[0]{%
Briceno:2014uqa,
Briceno:2015csa}

\newcommand{\eigenvaluePositive}[0]{%
Hansen:2012tf,
Briceno:2018mlh,
\dwave}

\newacronym{CMF}{CMF}{center-of-momentum frame}
\newcommand{\CMF}[0]{\gls{CMF}}

\title{Decay amplitudes to three hadrons from finite-volume matrix elements}
\author[1]{Maxwell T. Hansen}
\affiliation[1]{Higgs Centre for Theoretical Physics, School of Physics and Astronomy, The University of Edinburgh, Edinburgh EH9 3FD, UK}
\author[2]{, Fernando Romero-L\'opez}
\affiliation[2]{IFIC, CSIC-Universitat de Val\`encia, 46980 Paterna, Spain}
\author[3]{, and Stephen R. Sharpe}
\affiliation[3]{Physics Department, University of Washington, Seattle, WA 98195-1560, USA}

\emailAdd{maxwell.hansen@ed.ac.uk}
\emailAdd{fernando.romero@uv.es}
\emailAdd{srsharpe@uw.edu}

\abstract{
We derive relations between finite-volume matrix elements and infinite-volume decay amplitudes, for processes with three spinless, degenerate and either identical or non-identical particles in the final state. This generalizes the Lellouch-L\"uscher relation for two-particle decays and provides a strategy for extracting three-hadron decay amplitudes using lattice QCD. Unlike for two particles, even in the simplest approximation, one must solve integral equations to obtain the physical decay amplitude, a consequence of the nontrivial finite-state interactions. We first derive the result in a simplified theory with three identical particles, and then present the generalizations needed to study phenomenologically relevant three-pion decays. The specific processes we discuss are the CP-violating $K \to 3\pi$ weak decay, the isospin-breaking $\eta \to 3\pi$ QCD transition, and the electromagnetic $\gamma^*\to 3\pi$ amplitudes that enter the calculation of the hadronic vacuum polarization contribution to muonic $g-2$.
}

\allowdisplaybreaks
\begin{document}

\maketitle
\flushbottom
\clearpage

\section{Introduction}
\label{sec:intro}

The theoretical formalism for extracting three-hadron scattering amplitudes using lattice QCD has grown apace in recent years \cite{\ThreeBody}, and applications to simple systems have been successfully undertaken \cite{\ThreeQCDNumerics,Romero-Lopez:2018rcb,Romero-Lopez:2020rdq}. In all such studies, the basic approach is to extract the spectrum of three-hadron states in a finite spatial volume, and to use this information, by means of general relations, to constrain the infinite-volume scattering amplitudes. In particular, the spectrum of three-pion and three-kaon states of maximal isospin has been obtained in multiple calculations with different geometries, and with many values of total momentum in the finite-volume frame. In the following we abbreviate the latter as ``different frames''.

A natural extension of this work is to consider electroweak transitions to three particles, e.g.~the $K\to3 \pi$ decay. Although challenging, one can now conceive of undertaking a lattice calculation of finite-volume matrix elements of the form $\langle 3\pi, L |\mathcal H_W|K, L \rangle$, where $\mathcal H_W$ is the weak Hamiltonian density, and $\langle 3\pi, L \vert$ is a finite-volume state whose energy and momentum are tuned to match that of the initial kaon. Here we restrict attention to a cubic, periodic spatial volume, and $L$ denotes the periodicity (i.e.~the box length) in each of the three spatial dimensions. The question is then how to convert knowledge of several such matrix elements (with different volumes and frames) into information on the corresponding infinite-volume decay amplitude, including its dependence on the momenta of the three outgoing pions. In this work we answer this question, providing the formalism for a first-principles calculation of the amplitudes for $K\to 3\pi$ and related decays.

The corresponding problem for two-particle $K\to\pi\pi$ decays was solved in a seminal paper by Lellouch and L\"uscher (LL)~\cite{\LL}, where it was shown, for the case of a kaon at rest in the finite-volume frame, that the relation between the squared finite-volume matrix element and the magnitude squared of the infinite-volume decay amplitude is an overall multiplicative factor, the LL factor. This result was subsequently generalized in many ways~\cite{\transitions}, with the most important extension for our purposes being the work of refs.~\cite{\BHW,\BHLL}, in which an alternative and more general formalism was developed for calculating the LL factors for arbitrary $1\to 2$ processes mediated by an external operator. It is this approach that we use in the main text below to determine the generalization to three-particle final states.

To derive this generalization we first consider a final state consisting of three identical particles, and then move to the more phenomenologically interesting case of three pions in isosymmetric QCD. Exactly as in the two-particle case, the relation between finite-volume matrix elements and decay amplitudes follows from a quantization condition, which can be understood as a relation between finite-volume energies and hadronic scattering amplitudes. In this article we use the form of the quantization condition derived by two of us in refs.~\cite{\HSQCa,\HSQCb} together with its extension to all possible three pion states, derived by all of us in ref.~\cite{\isospin}. We refer to this approach as the relativistic field theory method.

We note, as was already stressed in refs.~\cite{\BHW,\BHLL}, that the basic methodology of relating finite-volume matrix elements to infinite-volume amplitudes can be applied to a wide range of processes. To emphasize this in the context of three-hadron final states, in this work we also describe in some detail how the approach may be applied to the virtual photon decay $\gamma^* \to 3 \pi$ as well as the isospin breaking transition $\eta \to 3 \pi$. The former process is relevant for quantifying finite-volume corrections to the hadronic-vacuum-polarization contribution to $(g-2)_\mu$ arising from the isoscalar part of the photon, along the same lines that $\gamma^* \to \pi \pi$ is used for the isovector part as described in refs.~\cite{\HVPfinitevolume}.

The remainder of the paper is organized into two parts. In the first, contained in section~\ref{sec:derivation}, we derive the necessary formalism for decays to states containing three identical particles. To do so, we first summarize the three-particle scattering formalism in section~\ref{sec:recap}. Then, in section~\ref{subsec:residual}, we derive the relation between the finite-volume matrix elements and a scheme-dependent intermediate infinite-volume quantity, $A^\text{PV}_{K3\pi}$. In section~\ref{subsec:threshold}, we explain how to systematically expand $A^\text{PV}_{K3\pi}$ about threshold based on symmetries, following which we explain how $A^\text{PV}_{K3\pi}$ can be connected to the physical decay amplitude via integral equations (section~\ref{subsec:PVtodecay}). To conclude the discussion for identical particles, in section~\ref{subsec:isotropic} we consider the isotropic approximation in which a more explicit and much simpler expression can be given, results from which we illustrate with numerical examples.

The second part of the paper, contained in section~\ref{sec:processes}, concerns the case of decays to three pions in isosymmetric QCD. We begin, in section~\ref{sec:physicalK3pi}, by presenting the appropriate generalization of the formalism. We then consider the processes $\gamma^* \to 3\pi$, $\eta \to 3\pi$ and $K^+ \to 3\pi$ in sections~\ref{subsec:gammato3pi}, \ref{subsec:etato3pi} and \ref{subsec:kto3pi}, respectively. We present our conclusions and outlook in section~\ref{sec:conc}.

We included four appendices. Appendix~\ref{app:A3pISA3dag} derives a technical result needed in the main text. Appendix~\ref{app:LLderivation} presents an alternative derivation of the relation obtained in section~\ref{subsec:residual} using the method of Lellouch and L\"uscher. Appendix~\ref{app:3pi} collects relevant results concerning the isospin decomposition of three-pion states. Finally, appendix~\ref{app:K0} presents the generalization of the results of section~\ref{subsec:kto3pi} to the decays of neutral kaons.

While this work was in preparation, a formalism for determining three-particle decay amplitudes to identical scalars in non-relativistic effective field theory (NREFT) was made public~\cite{Muller:2020wjo}. The authors considered only leading-order (non-derivative) couplings for the decay and scattering vertices. The formalism presented here goes beyond that of ref.~\cite{Muller:2020wjo} in several ways: (i) it is valid for nonidentical particles, and thus for the three-pion system; (ii) no approximations concerning the couplings are made, and no truncation in angular momenta is required; (iii) it is valid for generic moving frames; (iv) it is derived in a fully relativistic formalism.
We include additional brief comments on the relationship between the approaches in section~\ref{subsec:isotropic}.

\section{Derivation for identical particles}
\label{sec:derivation}

We consider first a simple theory consisting of two real scalar fields, the ``kaon" $K$ and ``pion'' $\phi$, both having an associated $\mathbb Z_2$ symmetry that conserves particle number modulo 2. Aside from this symmetry constraint, the interactions between these fields are arbitrary. The physical masses of the particles are $m_K$ and $m_\pi$, respectively, and satisfy
\begin{equation}
3 m_\pi < m_K < 5 m_\pi\,.
\label{eq:range}
\end{equation}
Both the kaon and the pion are stable particles in this theory. To induce decays, we add an interaction Hamiltonian, suggestively denoted $\mathcal H_W$, that violates both $\mathbb Z_2$ symmetries, and is chosen to couple the kaon to the odd-pion-number sector. A simple example of the required Hamiltonian density is
\begin{equation}
\mathcal H_W(x)=c_W \frac{K(x)\phi(x)^3}{3!}\,,
\label{eq:HWsimple}
\end{equation}
but we need not commit to a particular form; all that matters is that the interaction is local and has the correct quantum numbers. We treat $c_W$ as small, such that we need only work to first order in this parameter. Decays of the kaon to even numbers of pions, although kinematically allowed for two pions and possibly also for four pions, are forbidden by symmetries. The potential decay $K\to 5\pi$ is kinematically disallowed for the mass range in eq.~(\ref{eq:range}).

\begin{figure}
\includegraphics[width=\textwidth]{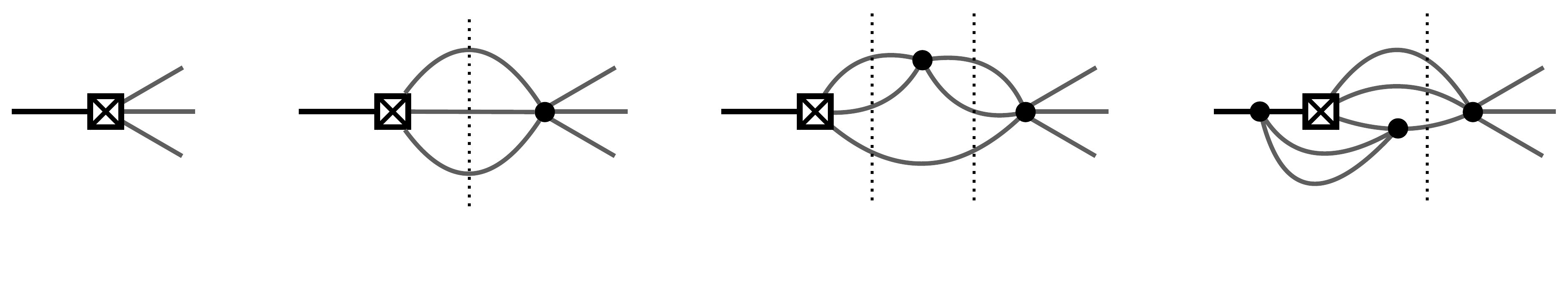}
\vspace{-40pt}
\caption{Examples of the underlying diagrams describing the $K \to 3 \pi$ decay and the corresponding finite-volume matrix element. The left-most diagram represents a local one-to-three transition with only exponentially suppressed finite-volume effects. By contrast the middle two diagrams have power-like $L$ dependence due to the on-shell intermediate states, indicated by the vertical dashed line. Finally, the rightmost diagram indicates a strong-interaction induced dressing to the weak vertex. All such interactions, as well as all dressing on the incoming and outgoing vertices are included in the formalism.\label{fig:diagrams}}
\end{figure}

To understand the intuition behind the following analysis, consider a diagrammatic representation of the $K \to 3 \pi$ amplitude, to leading order in $c_W$ but to all orders in the $\mathbb Z_2$ preserving interactions. As we illustrate in figure~\ref{fig:diagrams}, in such an expansion, the only on-shell intermediate states are those involving three pions. Arbitrary virtual interactions between the incoming (dressed) kaon and the final-state pions are allowed, but do not lead to on-shell intermediate states. One can think of such virtual loops as resulting from propagation that is localized near $\mathcal H_W$, and they lead to an effective renormalization of the bare coupling $c_W$. This is the physics that one expects to be captured by a calculation of the matrix element in a finite volume. On the other hand, the final-state interactions, which involve long distance, near on-shell propagation, will be mangled in finite volume, and it is these distortions that are corrected by the formalism developed in this work.

As stressed in the introduction, throughout this article we take finite volume to mean a cubic box of side $L$ with periodic boundary conditions on the fields $K$ and $\phi$. This restricts momenta to lie in the finite-volume set ${\bm p} = \bm n (2\pi/L)$, where $\bm n$ is a three-vector of integers. In our derivation, we drop volume-dependent terms that fall as $\exp(- m_\pi L)$ or faster. For typical volumes used in actual simulations, these exponentially-suppressed terms are much smaller than the power-law volume dependence that we keep. As is quite standard in these types of analyses, we take the temporal extent to be infinite. We also work in a continuum effective field theory with the assumption that the discretization effects entering a numerical lattice QCD calculation using these methods are small and included in the systematic uncertainties of the finite-volume matrix elements and energies.

\subsection{Recap of three-particle quantization condition and related formalism}
\label{sec:recap}

We make extensive use of the formalism developed to relate the finite-volume spectrum of three-particle states to the infinite-volume two- and three-particle scattering amplitudes. A general feature of the formalism is that it involves two steps. In the first, the finite-volume spectrum is related to an intermediate, unphysical infinite-volume three-particle K matrix ($\cK_{\df,3}$ in the approach of this paper), while, in the second, the K matrix is related to the scattering amplitudes by solving integral equations. This two-step procedure carries over naturally to the extension we develop here, with an intermediate, unphysical decay amplitude ($A_{K3\pi}^\PV$ below) determined from the finite-volume matrix elements, and the physical decay amplitude then obtained from $A_{K3\pi}^\PV$ via integral equations.

As noted above, we use the approach developed in refs.~\cite{\HSQCa,\HSQCb}, and our aim in this subsection is to recall its essential results. One important feature of this formalism for the case of identical particles is that the intermediate three-particle K matrix, $\cK_{\df,3}$, is symmetric under separate interchanges of initial and final momenta. This symmetry will carry over to the intermediate one-to-three amplitude, $A_{K3\pi}^\PV$, that arises here.\footnote{%
It is also possible to derive a simpler (though equivalent) version of the three-particle formalism that involves an asymmetric K matrix \cite{\BSQC} or the asymmetric R matrix \cite{\BSequiv}. We do not use these results, however, as the resulting renormalized decay amplitude is less constrained by symmetry, leading to a more complicated parametrization.}

The central result of ref.~\cite{\HSQCa} concerns the following three-particle finite-volume correlator:
\begin{equation}
C_L^{\sf M}(E,{\bm P}) = \int_{-\infty}^{\infty} dx_0 \int_L d^3x \, e^{i(E x^0 -\bm P\cdot \bm x)}
\langle0 | \text T \sigma(x) \sigma^\dagger(0)|0 \rangle\,,
\label{eq:CL3}
\end{equation}
where the superscript indicates that the underlying correlation function is evaluated in Minkowski space, and $\text T$ stands for time-ordering. Here $\sigma\sim \phi^3$ couples to three pions, but is otherwise an arbitrary operator possibly containing derivatives. Assuming the $\mathbb Z_2$ symmetry described above, the kinematic range of interest is
\begin{equation}
m_\pi < E^*=\sqrt{E^2-\bm P^2} < 5 m_\pi\,.
\label{eq:range2}
\end{equation}
Within this range, it is shown in ref.~\cite{\HSQCa} that the difference between the finite- and infinite-volume versions of this correlator takes the form\footnote{%
We are following the notation of ref.~\cite{\isospin} since we use results from this work in the physical $K\to3\pi$ case below. The notation differs slightly from that of refs.~\cite{\HSQCa,\HSQCb}.}
\begin{equation}
C_L^{\sf M}(E,\bm P) - C_\infty^{\sf M}(E,\bm P)
= i A_3'\frac1{F_3^{-1} + \cK_{\df,3}} A_3\,.
\label{eq:CL3res}
\end{equation}
Here all quantities have matrix indices $\{k\ell m\}$, with $A_3'$ a row vector, $A_3$ a column vector, while $F_3$ and $\cK_{\df,3}$ are matrices. The index $k$ is shorthand for the momentum $\bm k$ of one of the three particles, referred to as the spectator. The values of this index are drawn from the finite-volume set. The indices $\ell m$ give the decomposition into spherical harmonics of the angular dependence of the nonspectator pair, when boosted to the pair \CMF. The sum over $\bm k$ is cut off by a smooth function contained in $F$ and $G$, while the sum over $\ell$ is not cut off at this stage. All quantities are also implicit functions of $E$ and $\bm P$, with $F_3$ also depending on $L$. $F_3$ is given by
\begin{equation}
\label{eq:F3def}
F_3 = \frac1{2\omega L^3} \left[
\frac{F}3 - F \frac1{1+ \cM_{2,L} G} \cM_{2,L} F\right]\,,
\quad
\cM_{2,L}^{-1} = \cK_2^{-1} + F\,,
\end{equation}
where $\omega$, $F$, $G$, and $\cK_2$ are matrices defined in ref.~\cite{\HSQCa}, and (with the exception of $\omega$) are also implicit functions of $E$, $\bm P$ and, in the case of $F$ and $G$, also $L$. The only detail we need to know now is that $F$, $G$ and $\cK_2$ pick out one of the three particles as the spectator, so that these are intrinsically asymmetric quantities, an asymmetry that is inherited by $F_3$. By contrast, the endcaps $A'_3$ and $A_3$, as well as $\cK_{\df,3}$, are intrinsically symmetric quantities that are being expressed in terms of asymmetric variables.

The endcaps play an important role in the determination of the decay amplitude, as we will see below. The derivation of ref.~\cite{\HSQCa} defines these quantities by an all-orders constructive procedure, the key feature of which is that it involves loop integrals regulated by a principal value (PV) scheme. Thus one can think of the endcaps as, roughly speaking, the sum of all vacuum to three-pion diagrams in which only the short distance contributions from loops are kept. The long distance part, which leads to final state interactions, and the associated complex phases, is removed by the use of the PV prescription. We stress, however, that this qualitative interpretation of the endcaps is not needed to carry through the derivation described below. A technical result that is important below is that, if the creation and annihilation operators in $C_L^{\sf M}$ are related by hermitian conjugation, then $A_3' = A_3^\dagger$. We prove this fact in appendix \ref{app:A3pISA3dag}.

From the result (\ref{eq:CL3res}) for the correlator, the quantization condition is seen to be
\begin{equation}
\det (F_3^{-1} + \cK_{\df,3} ) = 0\,.
\label{eq:QC3}
\end{equation}
As written here, this equation ignores the residual symmetries of the finite-volume system that can be used to block diagonalize the matrix $F_3^{-1} + \cK_{\df,3}$. The relevant symmetry group depends on the value of $\boldsymbol P$. For the purposes of this work it suffices to note that for each group one can identify a set of irreducible representations (irreps), denoted by $\Lambda$, and for each irrep a row index, denoted $\mu$. Each set of $\Lambda \mu$ then corresponds to a block so that eq.~\eqref{eq:QC3} breaks into a set of independent quantization conditions of the form
\begin{equation}
\det_{\Lambda \mu} \big [ \mathbb P_{\Lambda \mu} \cdot ( F_3^{-1} + \cK_{\df,3} ) \cdot \mathbb P_{\Lambda \mu} \big ] = 0 \,,
\end{equation}
where $\mathbb P_{\Lambda \mu}$ projects out a given irrep and row.

To give the definition of $\mathbb P_{\Lambda \mu}$, we introduce $\mathbb R$ as a unitary matrix with the property that
\begin{equation}
\mathbb R^\dagger \cdot ( F_3^{-1} + \cK_{\df,3} ) \cdot \mathbb R \,,
\end{equation}
is block diagonal with one block corresponding to each possible value of $\Lambda \mu$. The construction of this matrix is a standard group-theoretic exercise, described, for example, in ref.~\cite{\dwave}. We then define $\widetilde {\mathbb P}_{\Lambda \mu}$ as a diagonal matrix of ones and zeroes that annihilates all blocks besides that corresponding to the target irrep and row. Finally we define
\begin{equation}
\label{eq:Pdef}
{\mathbb P}_{\Lambda \mu} = \mathbb R \cdot \widetilde {\mathbb P}_{\Lambda \mu} \cdot \mathbb R^\dagger \,,
\end{equation}
which projects to the target irrep while preserving the $\{k \ell m\}$ matrix space. The matrix $\mathbb P_{\Lambda \mu} \cdot ( F_3^{-1} + \cK_{\df,3} ) \cdot \mathbb P_{\Lambda \mu}$ will always have vanishing determinant, since the projection amounts to setting all eigenvalues with eigenvectors outside the $\Lambda \mu$ subspace to zero. For this reason, we include the $\Lambda \mu$ subscript on the determinant, indicating that this is evaluated only over the nontrivial subspace.

We stress that eqs.~\eqref{eq:QC3}-\eqref{eq:Pdef} are formal relations involving infinite-dimensional matrices and must be truncated in practice. This is done by assuming that the two- and three-particle interactions vanish above some value of $\ell$. For a given $\bm P$, $\Lambda \mu$ and $L$, this equation will be satisfied for a discrete set of values of $E$, which we label $E^{\Lambda}_n(\bm P,L)$ and often abbreviate as $E_n$.

The final result we need concerns the finite-volume three-particle scattering amplitude, $\cM_{3,L}$, defined in ref.~\cite{\HSQCb}. This is the finite-volume version of the amputated, connected infinite-volume amplitude $\cM_3$. What will be important here is how $\cM_{3,L}$ can be obtained from $C_L$ by an amputation procedure discussed in refs.~\cite{\HSQCb,\BHSK}. The idea is that, as we move in from the endcaps we may encounter a factor of $F$, and this sets the three particles on shell. An unsymmetrized form of the scattering amplitude, $\cM_{3,L}^{(u,u)}$, is then obtained by keeping terms in $C_L$ that have at least two factors of $F$---one for incoming and the other for outgoing particles---and dropping all but the contributions between the two outermost $F$s. In fact, this includes some disconnected three-particle diagrams that must also be dropped. In a final step, the resulting connected amplitude is symmetrized.

We now explain the resulting procedure in detail. We first remove the factors of $i$, $A_3'$ and $A_3$, and rewrite the result as\footnote{%
We remove the $i$ since the result of removing $A_3'$ and $A_3$ alone is $i \cM_{3,L}$.
}
\begin{align}
\frac1{F_3^{-1}+\cK_{\df,3}} &= F_3 - F_3\frac1{1+\cK_{\df,3} F_3} \cK_{\df,3} F_3\,, \label{eq:F3usefulrel}
\\
&= \frac{F}{6\omega L^3} - \frac{F}{2\omega L^3} \frac1{1+\cM_{2,L} G} \cM_{2,L} 2\omega L^3 \frac{F}{2\omega L^3}
- F_3\frac1{1+\cK_{\df,3} F_3} \cK_{\df,3} F_3\,.
\end{align}
We drop the first term on the right-hand side as it contains a single $F$, and complete the amputation by multiplying by the inverse of $iF/(2\omega L^3)$ on both ends. This leads to
\begin{equation}
\frac1{1+\cM_{2,L} G} \cM_{2,L} 2\omega L^3
+ \left(\frac{F}{2\omega L^3}\right)^{-1}
F_3\frac1{1+\cK_{\df,3} F_3} \cK_{\df,3} F_3
\left(\frac{F}{2\omega L^3}\right)^{-1}\,.
\end{equation}
Expanding out the first term in a geometric series, the leading contribution, $\cM_{2,L} 2\omega L^3$, is disconnected and thus dropped, leading to the final result for $\cM_{3,L}^{(u,u)}$,
\begin{align}
\cM_{3,L}^{(u,u)} &= \cD^{(u,u)} + \cL^{(u)}_L \frac1{1+\cK_{\df,3} F_3} \cK_{\df,3} \cR^{(u)}_L\,,
\label{eq:M3Luu}
\\
\cD^{(u,u)} &= - \frac1{1+\cM_{2,L} G} \cM_{2,L} G \cM_{2,L} 2\omega L^3 \,,
\label{eq:Duu}
\\
\cL^{(u)}_L &= \left(\frac{F}{2\omega L^3}\right)^{-1} F_3
= \frac13 - \frac1{1+\cM_{2,L} G} \cM_{2,L} F\,,
\label{eq:LLu}
\\
\cR^{(u)}_L &= F_3\left(\frac{F}{2\omega L^3}\right)^{-1}
= \frac13 - F \cM_{2,L}\frac1{1+G \cM_{2,L} }\,.
\label{eq:RLu}
\end{align}
The full amplitude is then given by
\begin{equation}
\cM_{3,L} = \cS\left\{ \cM_{3,L}^{(u,u)}\right\}\,,
\end{equation}
where the symmetrization operator is defined in ref.~\cite{\HSQCb}, and discussed in more detail in ref.~\cite{\isospin}. We also note that, following ref.~\cite{\HSQCb}, $\cM_3$ can be obtained from $\cM_{3,L}$ by taking the $L\to\infty$ limit in which poles in $F$ and $G$ are shifted from the real axis by the usual $i\epsilon$ prescription.

\subsection{Residue method to obtain intermediate decay matrix elements}
\label{subsec:residual}

The approach we follow is adapted from that of ref.~\cite{\BHLL}, and also draws from ref.~\cite{\BHW}. The matrix elements that can be determined in finite volume are
\begin{equation}
\langle E_n,\bm P, \Lambda \mu, L | \cH_W(0) | K, \bm P,L\rangle\,.
\label{eq:latME}
\end{equation}
Here $| K,\bm P,L\rangle$ is a single kaon state, with momentum $\bm P$ drawn from the finite-volume set, while $|E_n,\bm P, \Lambda \mu, L\rangle$ is a three-particle finite-volume state with the same momentum $\bm P$, and with energy $E_n$. It transforms in the irrep $\Lambda$ and in the row $\mu$ of that irrep. Both states are normalized to unity. The energy of the kaon state is $ E_K(\bm P) = (\bm P^2+m_K^2)^{1/2}$, with no volume dependence aside from exponentially suppressed effects. The energy of the three-particle state, by contrast, has a power-law dependence on $L$. In order to obtain a matrix element related to the infinite volume decay amplitude, $L$ should be tuned so that $E^{\Lambda}_n(\bm P, L)=E_K(\bm P)$, implying that four-momentum is conserved.\footnote{%
If one were interested in the matrix element (\ref{eq:latME}) in which $\cH_W(0)$ inserted energy, then the subsequent derivation would still hold in an appropriate kinematic regime. The analysis can also be straightforwardly generalized to the case where $\cH_W(0)$ inserts momentum.}
There can be many such matrix elements, each corresponding to a different finite-volume level, with a different choice of $L$ needed in each case.

It is useful to sketch how the matrix elements (\ref{eq:latME}) would be determined from a simulation of the theory, carried out necessarily in Euclidean space. We idealize the setup by assuming an infinite Euclidean time direction, and work with correlators fully transformed to momentum space. The three correlators that are needed are
\begin{align}
C_{K,L}(P) &= Z_K\int_{-\infty}^\infty dx_4\int_L d^3x\, e^{-i P x} \langle 0 \vert \text{T}_{\sf E} K(x_4,\bm x) K(0) \vert 0 \rangle\,,
\label{eq:CK}
\\
C_{3\pi,L}(P) &= \int_{-\infty}^{\infty} dx_4 \int_L d^3x\, e^{-iP x} \langle 0 \vert \text{T}_{\sf E} \cA_{3\pi}(x_4,\bm x) \cA_{3\pi}^\dagger(0) \vert 0 \rangle\,,
\label{eq:C3pi}
\\
C_{K3\pi,L}(P) &= \int_{-\infty}^\infty dx_4 \int_L d^3x\, e^{-iPx} \langle 0 \vert \text{T}_{\sf E} \cA_{3\pi}(x_4,\bm x) \cB_{K3\pi}(0) \vert 0 \rangle\,,
\label{eq:CK3pi}
\end{align}
where $P=(\bm P, P_4)$ and $x=(\bm x,x_4)$ are Euclidean four-vectors, whose inner product is denoted $Px$, and $\text T_{\sf E}$ denotes Euclidean time ordering.

The correlator $C_{K,L}$ determines the normalization constant $Z_K$. It should be chosen so that
\begin{equation}
\lim_{P_4\to i E_K(\bm P)} (P^2 + m_K^2) C_{K,L}(P) = 1\,,
\end{equation}
which implies that the renormalized kaon field satisfies
\begin{equation}
|\langle K, \bm P, L| \sqrt{Z_K} K(0) |0\rangle | = \frac1{\sqrt{2 E_K(\bm P) L^3}}\,.
\end{equation}

The correlator $C_{3\pi,L}$ determines the coupling of the operator $\cA_{3\pi}$ to the finite-volume states $|E_n, \bm P, \Lambda \mu,L\rangle$. Here, $\cA_{3\pi}$ is an operator chosen to couple to three-pion states in a particular row of the desired finite-volume irrep. In practice, $\cA_{3\pi}$ will involve pion fields with phase factors such that they have appropriate relative momenta, and thus will be complex. Other details of the operator are not relevant in the following. The correlator will consist of a sum of poles, and we pick out the contribution of the desired state from the residue
\begin{equation}
R_{3\pi}(E_n,\bm P,\Lambda \mu,L) \equiv \lim_{P_4\to i E_n} (E_n+i P_4) C_{3\pi,L}( P) = L^3 \left|
\langle 0 | \cA_{3\pi}(0) | E_n, \bm P, \Lambda \mu,L\rangle \right|^2\,.
\label{eq:C3pilim}
\end{equation}

The final correlator, $C_{K3\pi,L}$, can then be used to determine the desired matrix element. Here, following ref.~\cite{\BHLL}, we use a composite operator $\cB_{K3\pi}$ that both creates the initial kaon (implicitly having momentum $\bm P$) and includes the action of the weak Hamiltonian,
\begin{equation}
\cB_{K3\pi}(x) =\sqrt{ Z_K }
\lim_{P_4 \to i E_K(\bm P)} \left[ P^2 + m_K^2 \right] \int d^4 y\, e^{iP y} \cH_W(x) K(x+y) \,,
\label{eq:BK3pidef}
\end{equation}
where $P=(P_4,\bm P)$.
The limit picks out the incoming kaon pole, while the factor of $P^2+m_K^2$ amputates the kaon propagator.\footnote{Note that a subtlety arises here due to the fact that the operator $\cB_{K3\pi}$ is not local in time. This is not an issue because the $P_4 \to i E_K(\bm P)$ limit is dominated by early $y_4$ so that the $K(x+y)$ operator is ordered far to the right. Thus only one time-ordering arises, that with the intermediate finite-volume states that we analyze explicitly.}
Including all factors we obtain
\begin{align}
R_{K3\pi}(E_n,\bm P,\Lambda \mu,L) &\equiv
\lim_{P_4\to i E_n} (E_n+i P_4) C_{K3\pi,L}( P) \,,
\\
&\hspace{-70pt} = L^3
\langle 0 | \cA_{3\pi}(0) | E_n,\bm P,\Lambda \mu, L\rangle
\langle E_n, \bm P,\Lambda \mu, L| \cH_W(0)| K,\bm P,L \rangle \sqrt{2 E_K(\bm P) L^3} \,.
\label{eq:CK3pilim}
\end{align}
Without loss of generality, we can choose the phase of the operator and state such that $\langle 0 | \cA_{3\pi}(0) | E_n,\bm P, \Lambda \mu, L\rangle$ is real and positive. Then, combining eqs.~(\ref{eq:C3pilim}) and (\ref{eq:CK3pilim}), we obtain
\begin{equation}
\langle E_n, \bm P, \Lambda \mu, L| \cH_W(0)| K,\bm P,L \rangle \sqrt{2 E_K(\bm P) L^3} =
\frac{R_{K3\pi}(E_n,\bm P,\Lambda \mu,L)}{\sqrt{L^3 R_{3\pi}(E_n,\bm P,\Lambda \mu, L)}}
\,.
\label{eq:MEfinal}
\end{equation}
This matrix element will only be nonvanishing if $\Lambda$ and $\mu$ are chosen to match the transformation properties of $\cH_W(0) \vert K, \boldsymbol P, L \rangle$. If not, then the correlator $C_{K3\pi,L}(P)$ and the residue $R_{K3\pi}$ will vanish. For a rotationally invariant $\cH_W$, only the trivial irrep of the little group for momentum $\bm P$ will appear (or else the corresponding parity conjugate irrep), but we develop the formalism allowing for more general cases.

We now evaluate this ratio using the results from the previous subsection. To do so we first generalize the correlator $C_L$ of eq.~(\ref{eq:CL3}) by replacing $\sigma$ and $\sigma^\dagger$ with general operators $\cA$ and $\cB$ that couple the vacuum to three-pion states, but are, in general, unrelated to each other:
\begin{equation}
C_{AB,L}^{{\sf M}}(E,{\bm P}) = \int_{-\infty}^{\infty} dx_0 \int_L d^3x \, e^{i(E x^0 -\bm P\cdot \bm x)}
\langle0 | \text{T} \cA(x) \cB(0)|0 \rangle\,.
\label{eq:CABL}
\end{equation}
The analysis of ref.~\cite{\HSQCa} remains valid for $C_{AB,L}^{{\sf M}}$, since it requires only that the allowed on-shell intermediate states involve three pions. Thus the expression (\ref{eq:CL3res}) still holds, except that the endcaps $A_3'$ and $A_3$ are replaced by new quantities that we call, respectively, $A^\PV$ and $B^\PV$. The superscript is a reminder that loops in these quantities are defined using a PV prescription.

We next do a Wick rotation ($x_0\to -i x_4$) on the underlying correlation function, so that it is evaluated in Euclidean space-time. This results in
\begin{align}
C_{AB,L}^{{\sf M}}(E,\bm P) &= -i C_{AB,L}(P)\big|_{P_4=i E}\,,
\\
C_{AB,L}(P) &= \int_{-\infty}^\infty dx_4 \int_L d^3x\, e^{-i Px} \langle \text{T}_{\sf E} \cA(x) \cB(0) \rangle\,,
\label{eq:CABLdef}
\end{align}
where again $P=(\bm P,P_4)$.
It follows that $C_{AB,L}$ can be written
\begin{equation}
C_{AB,L}(P) = C_{AB,\infty}(P) - A^\PV \frac1{F_3^{-1} + \cK_{\df,3} }B^\PV\,,
\label{eq:CABLres}
\end{equation}
where now $A^\PV$, $F_3$, $\cK_{\df,3}$ and $B^\PV$ are written as functions of $P$ by setting $E=-iP_4$. The poles now lie on the imaginary axis, at the positions $P_4=i E_n$, where $E_n$ is a solution of the quantization condition eq.~(\ref{eq:QC3}).

The reason for these manipulations is that the two correlators that enter into the expression (\ref{eq:MEfinal}) for the desired matrix element, $C_{3\pi}$ and $C_{K3\pi}$, are in the class for which eq.~(\ref{eq:CABLres}) holds. In particular, we can use the results of ref.~\cite{\HSQCa} to write these correlators as
\begin{align}
C_{3\pi,L}(P) &= C_{3\pi,\infty}(P) - A_{3\pi}^\PV \frac1{F_3^{-1} + \cK_{\df,3} }A_{3\pi}^{\PV\dagger}\,,
\label{eq:C3piHS}
\\
C_{K3\pi,L}(P) &= C_{K3\pi,\infty}(P) - A_{3\pi}^\PV \frac1{F_3^{-1} + \cK_{\df,3} }A_{K3\pi}^\PV\,.
\label{eq:CK3piHS}
\end{align}
In eq.~(\ref{eq:C3piHS}) we are using the result, demonstrated in appendix \ref{app:A3pISA3dag}, that if the source and sink operators are related by hermitian conjugation, then the same holds for the endcap factors. Note that this only holds because the latter are defined with the PV prescription.

We next evaluate the residues that enter eq.~(\ref{eq:MEfinal}). Since the infinite-volume correlators and the endcaps are smooth, infinite-volume functions, $L$-dependent poles only arise from the zero eigenvalues in $F_3^{-1}+\cK_{\df,3}$. The required residues are thus
\begin{equation}
\cR_{\Lambda \mu}\big (E_n^\Lambda, \bm P, L \big ) = \lim_{P_4 \to i E_n^\Lambda} - (E_n^\Lambda + i P_4) \, \mathbb P_{\Lambda \mu} \cdot \frac{1}{F_3^{-1} + \cK_{\df,3}} \cdot \mathbb P_{\Lambda \mu},
\label{eq:residuedef}
\end{equation}
where the minus sign is for later convenience, and $E_n^{\Lambda}$ is one of the finite-volume three-pion energies for the given choice of $\bm P$, $\Lambda$ and $L$. $\cR_{\Lambda \mu}$ is a matrix in the $\{k \ell m\}$ space, which can be evaluated explicitly given expressions for $\cK_2$ (contained in $F_3$) and $\cK_{\df,3}$. The idea here is that these quantities have been previously determined (or, more realistically, constrained within some truncation scheme) by using the two- and three-particle quantization conditions applied to the spectrum of two- and three-particle states.

An important property of $\cR_{\Lambda \mu}$ is that it has rank one. This is because only one of the formally infinite tower of eigenvalues of $\mathbb P_{\Lambda \mu} \cdot (F_3^{-1} + \cK_{\df,3}) \cdot \mathbb P_{\Lambda \mu}$ will vanish for a given finite-volume energy $E_n^{\Lambda}(\bm P, L)$. Denoting the relevant eigenvalue by $\lambda(E,\bm P,\Lambda \mu,L)$ and the corresponding normalized eigenvector by $\boldsymbol e(E,\bm P,\Lambda \mu,L)$, one finds
\begin{equation}
\cR_{\Lambda \mu}\big (E_n^\Lambda, \bm P, L \big ) = \bigg ( \! \frac{\partial \lambda(E,\bm P,\Lambda \mu,L)}{\partial E} \bigg \vert_{E = E_n^{\Lambda}(\bm P, L)} \bigg )^{\!\!-1} \! \! \! \boldsymbol e(E,\bm P,\Lambda \mu,L) \, \boldsymbol e^\dagger(E,\bm P,\Lambda \mu,L) \,.
\end{equation}
This rank one property of $\mathcal R_{\Lambda \mu}$ was first described in the two-particle case in refs.~\cite{\BHW,\BHLL}. As is discussed, e.g.~in refs.~\cite{\eigenvaluePositive}, the eigenvalue must satisfy the inequality
\begin{equation}
\bigg ( \! \frac{\partial \lambda(E,\bm P,\Lambda \mu,L)}{\partial E} \bigg \vert_{E = E_n^{\Lambda}(\bm P, L)} \bigg )^{-1} > 0 \,,
\end{equation}
Thus, defining
\begin{equation}
v(E^{\Lambda}_n,\bm P,\Lambda \mu,L) \equiv \bigg ( \! \frac{\partial \lambda(E,\bm P,\Lambda \mu,L)}{\partial E} \bigg \vert_{E = E_n^{\Lambda}(\bm P, L)} \bigg )^{-1/2} \boldsymbol e(E,\bm P,\Lambda \mu,L) \,,
\end{equation}
$\cR_{\Lambda \mu}$ can be written as a simple outer product
\begin{equation}
\cR_{\Lambda \mu}(E^{\Lambda}_n,\bm P,L) = v(E^{\Lambda}_n,\bm P,\Lambda \mu,L) v^\dagger(E^{\Lambda}_n,\bm P,\Lambda \mu,L)\,.
\label{eq:residueres}
\end{equation}
Since $F_3^{-1}+\cK_{\df,3}$ is a real, symmetric matrix (assuming that we use real spherical harmonics), the elements of each $v$ are relatively real, with only the overall phase undetermined.

Using these results, we can immediately evaluate the required residues, obtaining
\begin{align}
R_{3\pi}(E^{\Lambda}_n,\bm P,\Lambda \mu,L) &= |A_{3\pi}^\PV v |^2\,,
\label{eq:R3piRFT}
\\
R_{K3\pi}(E^{\Lambda}_n,\bm P,\Lambda \mu,L) &= (A_{3\pi}^\PV v ) (v^\dagger A_{K3\pi}^\PV)\,,
\label{eq:RK3piRFT}
\end{align}
where $v$ is an abbreviation for $v(E^{\Lambda}_n,\boldsymbol P,\Lambda \mu,L)$. All quantities on the right-hand side are (implicitly) evaluated at $P=(\bm P, iE^{\Lambda}_n)$, with $E^{\Lambda}_n=E_K(\bm P)$. The overall sign in eq.~(\ref{eq:residuedef}) can now be justified. From eq.~(\ref{eq:C3pilim}), we know that $R_{3\pi}$ is positive, and thus the overall sign in eq.~(\ref{eq:R3piRFT}) must be positive, as shown.\footnote{%
This is in fact the criterion introduced in ref.~\cite{\BHSnum}, and studied in refs.~\cite{\dwave,\largera}, to determine whether solutions to the three-particle quantization condition are physical.
}

Choosing the phase of $v$ such that $A_{3\pi}^\PV v$ is real and positive, and inserting these results into eq.~(\ref{eq:MEfinal}), we obtain
\begin{equation}
\sqrt{2 E_K(\bm P) }L^3 \langle E_n, \bm P, \Lambda \mu,L |\cH_W(0)| K,\bm P,L \rangle =
v^\dagger A_{K3\pi}^\PV
\,.
\label{eq:MEfinal2}
\end{equation}
This achieves the aim of relating the finite-volume decay matrix element (which could be determined by a numerical simulation) to a quantity in the generic relativistic field theory, namely a projection of the quantity $A^\PV_{K3\pi}$. By using multiple matrix elements, one could determine the parameters in a truncated approximation to $A^\PV_{K3\pi}$. The result (\ref{eq:MEfinal2}) can also be derived by a generalization of the method of Lellouch and L\"uscher~\cite{\LL}, as we show in appendix \ref{app:LLderivation}.

Before turning to parametrizations of $A^\PV_{K3\pi}$, we close this section with a few more comments on the phase conventions entering the various relations on matrix elements. We first review the requirements we have imposed above. First, we have fixed the phase of the state $ \cA_{3\pi}(0) | E_n,\bm P, \Lambda \mu, L\rangle$ by requiring that $\langle 0 | \cA_{3\pi}(0) | E_n,\bm P, \Lambda \mu, L\rangle $ is real and positive. Second, we have required that, while $A_{3\pi}^\PV$ and $v(E^{\Lambda}_n,\bm P,\Lambda \mu,L)$ may individually carry phases, these must cancel such that $A_{3\pi}^\PV v$ is real and positive. We have then demonstrated that, with these two convention choices, the finite-volume matrix element appearing in eq.~\eqref{eq:MEfinal2} must have the same phase as the combination $v^\dagger A_{K 3\pi}^{\text{PV}}$. Finally, to extract the value of $A_{K 3\pi}$, we must establish the phase of $v$ itself, which has been left open so far. The most natural convention is to simply require $A_{3\pi}^\PV$ and $v$ to be individually real. In this convention $v^\dagger$ is also real, so any phase in the finite-volume matrix element on the left-hand side of eq.~(\ref{eq:MEfinal2}) (resulting, for example, from a CP-violating phase in $\cH_W$) will be inherited by $A_{K3\pi}^\PV$.

As was already discussed in refs.~\cite{\phaseRefs}, the utility in carefully tracking this phase information is that it allows one to extract relative phases between various matrix elements. For example, if the weak Hamiltonian density is decomposed into operators $\cO_1(x)$ and $\cO_2(x)$, it follows from eq.~(\ref{eq:MEfinal2}) that
\begin{equation}
\frac{\langle E_n, \bm P, \Lambda \mu,L |\mathcal O_1(0)| K,\bm P,L \rangle}{ \langle E_n, \bm P, \Lambda \mu,L |\mathcal O_2(0)| K,\bm P,L \rangle} = \frac{v^\dagger A_{K3\pi[\mathcal O_1]}^\PV }{v^\dagger A_{K3\pi [\mathcal O_2]}^\PV} \,.
\end{equation}
The overall phase in $v^\dagger$ cancels, so the phase in the ratio of PV amplitudes on the right-hand side is given by that of the ratio of the matrix elements on the left-hand side. This phase information will be passed on to the decay matrix elements by solving the integral equations described below in section \ref{subsec:PVtodecay}.

\subsection{Threshold expansion of $A^\PV_{K3\pi}$}
\label{subsec:threshold}

Since $A^\PV_{K3\pi}$ is an unfamiliar quantity, we discuss its properties in this brief subsection. We recall that it is an infinite-volume on-shell quantity, given, crudely speaking, by calculating all $K\to3\pi$ diagrams with PV regulation for the poles. Thus it is an analytic function of the kinematic variables, symmetric under interchange of any pair of final-state momenta.

A useful parametrization of $A^\PV_{K3\pi}$ is given by the threshold expansion, which is an expansion in powers of relativistic invariants that vanish at threshold, for instance
\begin{equation}
\Delta=\frac{m_K^2-9 m_\pi^2}{9 m_\pi^2}\,.
\end{equation}
For the decays $K^+\to\pi^+\pi^+\pi^-$ and $K^+\to \pi^+ \pi^0 \pi^0$, for example, $\Delta\approx 0.39$ and $0.45$, respectively. Labelling the pion four-momenta $p_1$, $p_2$, and $p_3$, so that $P=p_K=p_1+p_2+p_3$, the three Mandelstam variables are
\begin{equation}
s_i=(p_j+p_k)^2 = (P-p_i)^2\,, \quad \sum_{i=1}^3 s_i = m_K^2+ 3m_\pi^2\,,
\end{equation}
where $\{i,j,k\}$ are ordered cyclically. We will expand in dimensionless quantities that vanish at threshold, namely $\Delta$ and
\begin{equation}
\label{eq:DeltaiDef}
\Delta_i =\frac{s_i-4m_\pi^2}{9 m_\pi^2}\,,
\end{equation}
which satisfy $\sum_i \Delta_i = \Delta$. Using this sum rule, and enforcing particle-interchange symmetry and smoothness, we find\footnote{%
The presence of only a single term in each of the second, third and fourth orders is a pattern that does not continue to higher orders.}
\begin{align}
A^\PV_{K3\pi} &= A^{\sf iso}
+ A^{(2)} \sum_i \Delta_i^2
+ A^{(3)} \sum_i \Delta_i^3
+ A^{(4)} \sum_i \Delta_i^4
+ \cO(\Delta^5)\,.
\label{eq:APVexp}
\end{align}
Here ``iso'' refers to the isotropic limit, in which the amplitude is independent of the momenta of the decay products. To obtain a strict expansion in powers of $\Delta$, one would need to expand the coefficients, e.g.
\begin{equation}
A^{\sf iso} = \sum_{n=0}^\infty \Delta^n A^{\sf iso,n}
\,,
\end{equation}
keeping only the appropriate number of terms (e.g. the first five terms if working to fourth order in $\Delta$).

To use the threshold expansion (\ref{eq:APVexp}) in the result from the previous subsection, eq.~(\ref{eq:MEfinal2}), one must convert $A^\PV_{K3\pi}$ to the $\{k\ell m\}$ basis. We recall here how this is done~\cite{\HSQCa}. We first note that the on-shell three-particle phase space with fixed total four-momentum (and ignoring Lorentz invariance) is five-dimensional. We can parametrize this space in various ways, one choice being to use a set of five momentum coordinates: $p_{1,x}$, $p_{1,y}$, $p_{1,z}$, $p_{2,x}$, $p_{2,y}$. The remaining five coordinates are then set by the fixed total energy and momentum. To connect to the $\{k\ell m\}$ basis we make a different choice, labelled $\{ \bm k, \widehat{\bm a}^*\}$. Here $\bm k$ is one of the three momenta, e.g. $\bm k= \bm p_1$, while $\widehat{\bm a}^*$ is the result of boosting the remaining two particles to their \CMF~and picking the direction of one of them, say particle 2. Here we are using the notation that a quantity with a superscript $*$ is evaluated in a boosted frame. We then decompose the amplitude into spherical harmonics in the pair \CMF,
\begin{equation}
A^\PV_{K3\pi}(\bm k,\widehat{\bm a}^*) = \sum_{\ell m}
\sqrt{4\pi}\, Y_{\ell m}(\widehat{\bm a}^*) A^\PV_{K3\pi}(\bm k)_{\ell m}\,.
\label{eq:decompose}
\end{equation}
To use the result of the previous subsection we must restrict $\bm k$ to lie in the finite-volume set,
\begin{equation}
A^\PV_{K3\pi; k\ell m} \equiv A^\PV_{K3\pi}(\bm k)_{\ell m}\Big|_{\bm k=2\pi \bm n/L}\,.
\end{equation}

The decomposition of the terms in the threshold expansion into the $\{k\ell m\}$ basis is straightforward but tedious, and we do not present it here. It follows closely the corresponding decomposition of $\cK_{\df,3}$ worked out in ref.~\cite{\dwave}.

\subsection{Relating $A_{K3\pi}^\PV$ to the physical decay amplitude}
\label{subsec:PVtodecay}

In this subsection we show how the physical $K\to3 \pi$ decay amplitude can be obtained by solving appropriate integral equations, once the endcap $A_{K3\pi}^\PV$ has been determined using the results of the previous two subsections. This is the second step of the general procedure described in section~\ref{sec:recap}, and involves relations between infinite-volume quantities. The method we use follows the strategy introduced in ref.~\cite{\HSQCb}: we consider a finite-volume correlator whose infinite-volume limit produces the physical decay amplitude, and write this correlator in terms of $\cK_2$, $\mathcal K_{\text{df},3}$, and in particular $A_{K3\pi}^\PV$.

We begin by recalling that the infinite-volume decay matrix element can be defined by
\begin{equation}
\label{eq:TK3pidef}
T_{K3\pi} = \bra{3\pi,{\rm out}} \cH_W(0) \ket{K, \bm P}\,,
\end{equation}
where states are defined using the standard relativistic normalization. The decay rate is then given by
\begin{equation}
\Gamma =
\frac1{3!} \frac1{2m_K} \int
\textrm{dLIPS}\,
|T_{K3\pi}|^2 \,,
\label{eq:decayrate}
\end{equation}
where $1/3!$ is the identical-particle symmetry factor, and dLIPS is the Lorentz-invariant phase-space measure. We will use the $\{\bm k, \widehat{\bf a}^*\}$ variables introduced above, in terms of which the measure becomes
\begin{equation}
\textrm{dLIPS}
=
\frac{d^3 \bm k}{2\omega_k (2\pi)^3}
\frac{a^*}{4\pi \omega_{a^*}}
\frac{d^2\Omega_{\hat{\bm a}^*}}{4\pi}\,.
\end{equation}
Here $a^{*2} = q_{2,k}^{*2}$ is the squared momentum of one of the nonspectator pair in their \CMF, with
\begin{equation}
q_{2,k}^{*2} = (E_K(\bm P) - \omega_k)^2 - (\bm P - \bm k)^2\,,
\label{eq:q2kstar}
\end{equation}
and $\omega_{a^{*}}= \sqrt{a^{*2}+m_\pi^2}$ is the corresponding energy.

In order to obtain an expression for $T_{K3\pi}$ in terms of $A_{K3\pi}^\PV$, we consider the finite-volume decay matrix element, $T_{K3\pi,L}$. This is defined as the sum of all Feynman diagrams contributing to $T_{K3\pi}$, including appropriate amputations, but evaluated with finite-volume Feynman rules. A subtlety arises because the energies of three external on-shell pions, each with a momentum from the finite-volume set will, not, in general, sum to $E_K(\bm P)$. To have an energy-conserving process, the external momenta in $T_{K3\pi,L}$ must be adjusted. This on-shell projection is done using the method introduced in ref.~\cite{\HSQCa}. The spectator momentum, $\bm k$, is held fixed at a finite-volume value, while the magnitude of $\bm a^*$ (the momentum of one of the nonspectator pair boosted to the pair CMF) is adjusted until energy is conserved. This requires setting $\bm a^*=q_{2,k}^* \widehat{\bm a}^*$, and leads to the third particle having momentum $-\bm a^*$ in the pair \CMF. This is the on-shell projection that appears in all quantities adjacent to factors of $F$ and $G$. The projection only affects the external momenta for $T_{K3\pi,L}$---when written as a skeleton expansion in terms of Bethe-Salpeter kernels, the internal loop momenta are all drawn from the finite-volume set. This point is discussed at length in ref.~\cite{\HSQCb}. The result is the quantity $T_{K3\pi,L}(\bm k, \widehat{\bm a}^*)$.

We will need a variant of this quantity in the following, namely $T^{(u)}_{K3\pi,L}(\bm k, \widehat{\bm a}^*)$, which we refer to as the asymmetric decay amplitude. This is defined as the sum of the same set of amputated diagrams with two restrictions: First, if the final interaction involves a two-particle Bethe-Salpeter kernel, then $\bm k$ is chosen as the momentum for the spectator particle. Second, if the final interaction involves a three-particle kernel, then the diagram is multiplied by $1/3$. In fact, what appears in the expressions below is $T^{(u)}_{K3\pi,L,k\ell m}$, which results when we decompose the $\widehat{\bm a}^*$ dependence into spherical harmonics as in eq.~(\ref{eq:decompose}).

To obtain the desired expression for $T_{K3\pi,L,k\ell m}$, we begin from the correlator $C_{K3\pi,L}(P)$, introduced in eq.~(\ref{eq:CK3pi}), which describes a finite-volume $K\to3\pi$ process. We consider the Minkowski version of this correlator, given by
\begin{equation}
C_{K3\pi,L}^{{\sf M}}(E,\bm P) = C_{K3\pi,\infty}^{{\sf M}}(E,\bm P)
+ A_{3\pi}^\PV \frac{i}{F_3^{-1} + \cK_{\df,3} }A_{K3\pi}^\PV\,.
\label{eq:CK3piHSM}
\end{equation}
We obtain $T_{K3\pi,L}$ by keeping contributions that have at least one factor of $F$ (since this puts the intermediate three-particle state on shell) and amputating all that lies to the left of the left-most $F$. Only the second term on the right-hand side contains $F$s, and we amputate it as described in section~\ref{sec:recap} by removing $A_{3\pi}^\PV$ and multiplying by the inverse of $iF/(2\omega L^3)$, leading to
\begin{align}
T_{K3\pi,L}^{(u)} &=
\left(\frac{iF}{2\omega L^3}\right)^{-1} F_3 \frac{i}{1+\cK_{\df,3} F_3} A_{K3\pi}^\PV\,,
\\
&= \cL_L^{(u)} \frac1{1+\cK_{\df,3} F_3} A_{K3\pi}^\PV\,,
\label{eq:Tufinal}
\end{align}
where $\cL_L^{(u)}$ is given in eq.~(\ref{eq:LLu}). Note that, unlike in the construction of $\cM_{3,L}^{(u,u)}$ described in section~\ref{sec:recap}, here there are no disconnected terms to drop.

With the expression for $T^{(u)}_{K3\pi;k\ell m}$ in hand, we next note, following ref.~\cite{\HSQCb}, that the result can be extended to an arbitrary choice of $\bm k$, not just one in the finite-volume set. The form of eq.~(\ref{eq:Tufinal}) remains unchanged, and the various quantities extend simply to arbitrary $\bm k$, as explained in ref.~\cite{\HSQCb}. The result, $T^{(u)}_{K3\pi,L}(\bm k)_{\ell m}$, is still a finite-volume quantity, since internal loops remain summed. We now insert $i\epsilon$ factors to regulate the poles in $F$ and $G$, and take the infinite-volume limit holding $\bm k$ fixed
\begin{equation}
T^{(u)}_{K3\pi}(\bm k)_{\ell m} = \lim_{\epsilon\to 0^+} \lim_{L\to \infty}
T^{(u)}_{K3\pi,L}(\bm k)_{\ell m}\bigg|_{E \to E + i\epsilon}\,.
\label{eq:limTu}
\end{equation}
This gives the correct asymmetric infinite-volume decay amplitude because, in the limit, all sums in Feynman diagrams that run over a pole (which are those in which three particles can go on shell) are replaced by integrals in which the pole is regulated by the standard $i\epsilon$ prescription.

The final step is to obtain the complete decay amplitude by symmetrizing, which corresponds to adding all possible attachments of the momentum labels to the Feynman diagrams. This is effected by
\begin{align}
T_{K3\pi}(\bm k, \widehat{\bm a}^*) &\equiv \cS\left\{ T_{K3\pi}(\bm k)_{\ell m} \right\} \,,
\label{eq:symmetrize}
\\
&=
T^{(u)}_{K3\pi}(\bm k, \widehat{\bm a}^*)
+
T^{(u)}_{K3\pi}(\bm a, \widehat{\bm b}^*)
+
T^{(u)}_{K3\pi}(\bm b, \widehat{\bm k}^*)
\,,
\label{eq:symTu}
\end{align}
where $T^{(u)}_{K3\pi,L} (\bm k,\widehat{\bm a}^*)$ is obtained by combining $T^{(u)}_{K3\pi,L}(\bm k)_{\ell m}$ with spherical harmonics as in eq.~(\ref{eq:decompose}). The notation in eq.~(\ref{eq:symTu}) is the natural generalization of that given above: just as $(\omega_{a^*},\bm a^*)$ is the result of boosting $(\omega_a,\bm a)$ to the \CMF~of the $\{\bm a, \bm b\}$ pair (with $\bm b=\bm P-\bm k-\bm a$), so $(\omega_{b^*},\bm b^*)$ is the result of boosting $(\omega_b,\bm b)$ to the \CMF~of the $\{\bm b,\bm k\}$ pair, while $(\omega_{k^*},\bm k^*)$ is the result of boosting $(\omega_k,\bm k)$ to the \CMF~of the $\{\bm k,\bm a\}$ pair.

Applying this procedure to the result eq.~(\ref{eq:Tufinal}) for $T_{K3\pi,L}^{(u)}$ leads to a set of integral equations. Since the steps are very similar to those in ref.~\cite{\HSQCb}, we simply quote the final results. As for $T^{(u)}_{K3\pi}$, the $\{k\ell m\}$ indices used in finite volume go over in infinite-volume to a dependence on the continuous spectator momentum, $\bm k$, as well as an unchanged dependence on $\ell$ and $m$. Thus the matrix indices $\ell m$ remain, and will be implicit in the following equations, while the dependence on $\bm k$ will be explicit.

The combination $(1+\cM_{2,L} G)^{-1} \cM_{2,L}$, which appears in $\cL^{(u)}_L$ and in $F_3$, goes over in infinite volume to $\cD_{23}^{(u,u)}(\bm p,\bm k)_{\ell' m';\ell m}$ (using the notation of ref.~\cite{\BSQC}), which satisfies
\begin{equation}
\cD_{23}^{(u,u)}(\bm p,\bm k) = \overline{\delta}(\bm p-\bm k) \cM_2(\bm k) - \cM_2(\bm p) \int_{\bm r} G^\infty(\bm p, \bm r) \cD_{23}^{(u,u)}(\bm r, \bm k)
\,,
\label{eq:D23uu}
\end{equation}
where $G^\infty$ is defined in eq.~(81) of ref.~\cite{\HSQCb}, and includes an $i\epsilon$-regulated pole, while
\begin{align}
\overline \delta(\bm p-\bm k) &= 2\omega_p (2\pi)^3 \delta^3(\bm p-\bm k) \,,
\\
\cM_2(\bm k)_{\ell' m';\ell m} &= \delta_{\ell' \ell} \delta_{m' m} \cM_2^{(\ell)}(q_{2,k}^*)\,,
\\
\int_{\bm r} &= \int \frac{d^3r}{2\omega_r (2\pi)^3}\,.
\end{align}
Here $\cM_2^{(\ell)}$ is the $\ell$th partial wave of $\cM_2$, evaluated for the \CMF~momentum of one of the scattering pair. Given a solution to the integral equation (\ref{eq:D23uu}), and the relation of $F_3$ to $\cL^{(u)}_L$, eq.~(\ref{eq:LLu}), the equation satisfied by the infinite-volume limit of $X=(1 +\cK_{\df,3} F_3)^{-1}$ is
\begin{equation}
X(\bm p,\bm k) = \overline{\delta}(\bm p-\bm k) - \int_{\bm r, \bm s}
\cK_{\df,3}(\bm p, \bm r) \widetilde \rho_\PV(\bm r)
\cL^{(u)}(\bm r, \bm s) X(\bm s, \bm k)\,.
\label{eq:X}
\end{equation}
In the first term there is an implicit identity matrix in $\ell m$ space. The quantity $\widetilde \rho_\PV$ results from the infinite-volume limit of $F$, and is
\begin{equation}
\widetilde \rho_\PV(\bm r)_{\ell' m';\ell m} =
\delta_{\ell' \ell} \delta_{m' m}\; \widetilde \rho_\PV^{(\ell)}(q_{2,r}^*)
\,,
\label{eq:rhoPV}
\end{equation}
where $\rho_\PV^{(\ell)}$ is a modified phase space factor given in eq.~(B6) of ref.~\cite{\BSQC}.
Finally,
\begin{equation}
\cL^{(u)}(\bm r,\bm s) = \frac13 \overline\delta(\bm r-\bm s) - \cD_{23}^{(u,u)}(\bm r, \bm s)
\widetilde \rho_\PV(\bm s) \,,
\label{eq:Lu}
\end{equation}
which is the infinite-volume limit of $\cL^{(u)}_L$.

With these ingredients we can write down the relationship of the asymmetric decay amplitude to $A_{K3\pi}^\PV$,
\begin{equation}
T^{(u)}_{K3\pi}(\bm k) = \int_{\bm r, \bm s}
\cL^{(u)}(\bm k,\bm r) X(\bm r, \bm s) A_{K3\pi}^\PV(\bm s)\,.
\label{eq:Tures}
\end{equation}
The full amplitude is then given by symmetrization
\begin{equation}
T_{K3\pi}(\bm k,\widehat{\bm a}^*) = \cS\left\{
T^{(u)}_{K3\pi}(\bm k)_{\ell m} \right\}\,,
\label{eq:TK3pifinal}
\end{equation}
using the definition in eq.~(\ref{eq:symmetrize}) above. This completes the procedure for determining the decay amplitude from the finite-volume decay matrix elements. The physical interpretation of the factors in eq.~(\ref{eq:Tures}) is as follows. $\cL^{(u)}$ incorporates pairwise final state interactions, through multiple factors of $\cM_2$ alternating with switch factors $G^\infty$. $T^{(u)}_{K3\pi}$ becomes complex both because $\cM_2$ itself is complex, and due to the $i\epsilon$ in $G^\infty$. The quantity $X$ incorporates final state interactions involving all three particles, with intermediate pairwise scattering. Since this result derives from an all-orders diagrammatic derivation, the amplitude $T_{K3\pi}$ will automatically satisfy the required unitarity constraints, and in particular those that lead to Khuri-Treiman relations describing final-state interactions~\cite{\KT}.

\subsection{Isotropic approximation}
\label{subsec:isotropic}

We close this section by giving an explicit example of how the formalism works when making the simplest approximations to the decay and scattering amplitudes. We assume that only the leading, isotropic term in the threshold expansion of the decay amplitude, $A^{\sf iso}$, is nonvanishing---see eq.~(\ref{eq:APVexp}). This implies that $A^\PV_{K3\pi;k\ell m}$ is only nonzero for $\ell=m=0$, and is independent of $k$. In addition, it couples only to three-pion states in the trivial irrep of the appropriate little group, e.g., the $A^-_1$ irrep for $\bm P = 0$ (for pions with negative intrinsic parity). For the amplitudes $\cM_2$ and $\cK_{\df,3}$, we assume that only the $s$-wave contributes (so again $\ell=m=0$) and that $\Kdf$ is independent of the spectator momentum. This is equivalent to keeping only the isotropic term in the threshold expansion of $\Kdf$~\cite{\BHSnum,\dwave}.

\begin{figure}[tbh]
\centering\includegraphics[width=0.9\textwidth]{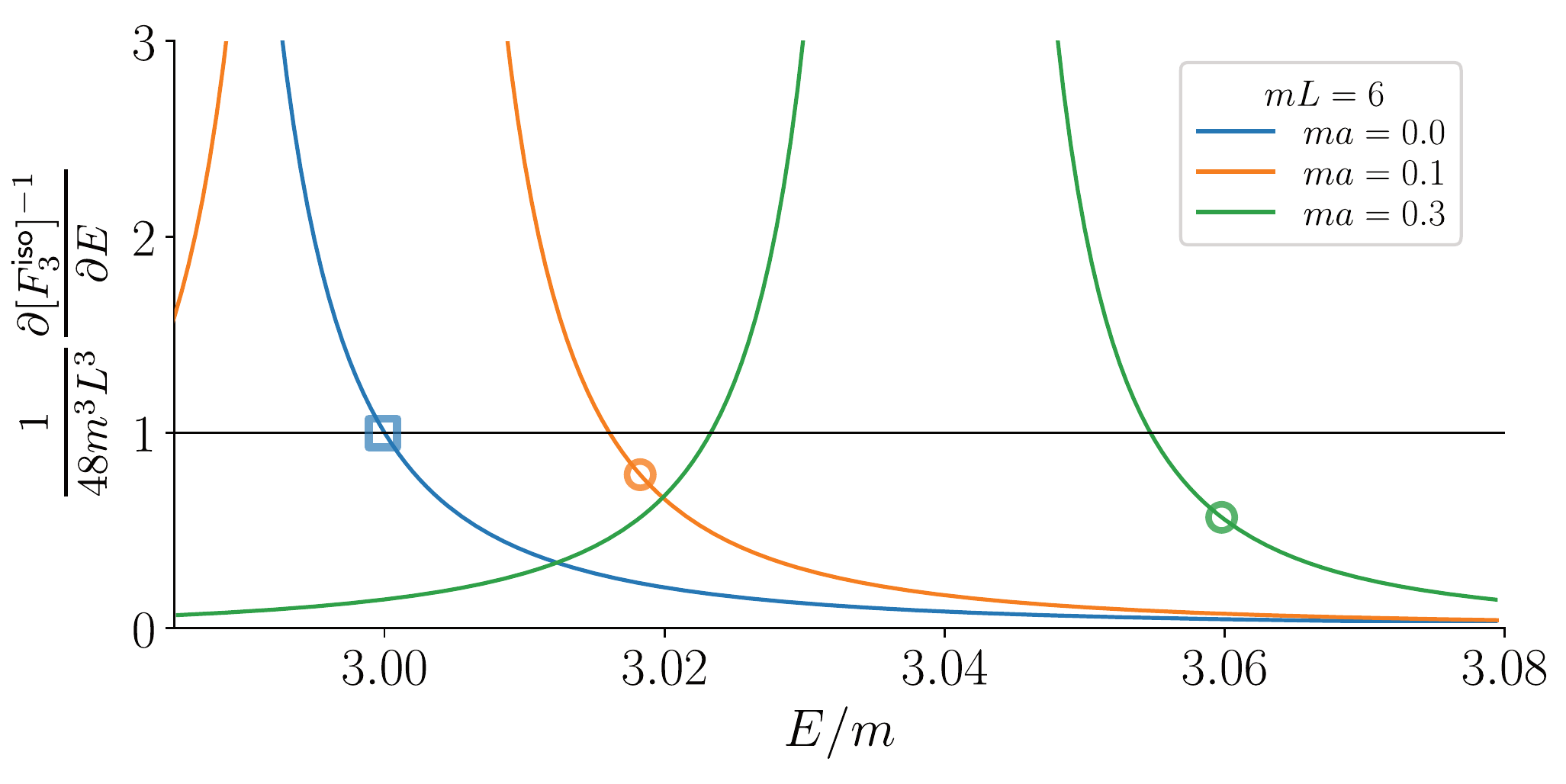}
\caption{ Plot of the conversion factor appearing in eq.~\eqref{eq:LLsquaredform} (rescaled as indicated by the plot label) in the vicinity of the three-particle threshold for the case of constant $\mathcal K^{\sf iso}_{\text{df},3}$. The factor is plotted versus energy $E$ for $\boldsymbol P = \boldsymbol 0$ and $mL=6$. The two-particle K matrix, entering $F_3^{\text{iso}}$, determined by keeping only the scattering length, $a$, in the effective range expansion. The three curves correspond to three values of the scattering length, as indicated by the legend, and each unfilled marker corresponds to the ground-state energy for the corresponding $ma$ value when $\mathcal K^{\sf iso}_{\text{df},3}=0$. In particular, the blue square corresponds to the non-interacting limit. The fact that the conversion factor is unity in the latter case indicates that the non-interacting matrix elements are equal in finite and infinite volume, up to a trivial normalization. More generally, once the scattering length is determined, these types of curves allow one to directly relate---within the isotropic approximation---any value of measured three-particle energy (horizontal axis) to a matrix element conversion factor (vertical axis). }
\label{fig:LLPlot}
\end{figure}

Given these approximations, all quantities entering the definition of $F_3$ depend only on the spectator momenta. The isotropic nature of $A^\PV_{K3\pi}$ and $\cK_{\df,3}$ is represented by introducing the vector $\ket{1}$ in spectator-momentum space, which equals unity for all choices of $\bm k$ in the finite-volume set that lie below the cutoff. Specifically,
\begin{equation}
A^\PV_{K3\pi} \longrightarrow \ket{1} A^{\sf iso}\ \ {\rm and}\ \
\cK_{\df,3} \longrightarrow \ket{1}\cK_{\df,3}^{\sf iso} \bra{1}
\,,
\end{equation}
where $A^{\sf iso}$ and $\cK^{\sf iso}_{\df,3}$ are constants. Using eq.~(\ref{eq:F3usefulrel}), one then finds that
\begin{align}
\frac1{F_3^{-1} + \cK_{\df,3}} &\longrightarrow
F_3 - F_3\ket{1} \frac1{F_3^{\sf iso} + (\cK_{\df,3}^{\sf iso})^{-1}} \bra1 F_3\,,
\label{eq:isoprop}
\end{align}
where $F_3^{\sf iso}$ is the isotropic component of $F_3$,
\begin{equation}
F_3^{\sf iso} \equiv \bra1 F_3 \ket 1\,.
\end{equation}
It follows that the only poles in three-particle correlators [e.g. $C_L^{\sf M}$ of eq.~(\ref{eq:CL3res})] that depend on $\cK_{\df,3}^{\sf iso}$ occur when the isotropic quantization condition is satisfied, i.e.
\begin{equation}
F_3^{\sf iso} = - (\cK_{\df,3}^{\sf iso})^{-1} \,.
\end{equation}
There are also solutions at free energies resulting from the $F_3$ terms in eq.~(\ref{eq:isoprop}), but these are an artifact of the isotropic approximation, as discussed in Appendix F of ref.~\cite{\dwave}. From eq.~(\ref{eq:isoprop}), we can determine the residue using eq.~(\ref{eq:residuedef}), finding
\begin{equation}
\cR^{\sf iso}_n = F_3\ket1 r_n^{\sf iso} \bra1 F_3\,,
\label{eq:Rnisores}
\end{equation}
where we have abbreviated the arguments of $\cR_{\Lambda \mu}(E_n, \boldsymbol P, L)$, and defined
\begin{equation}
r_n^{\sf iso} = - \bigg ( \frac{\partial F_3^{\sf iso} (E, \bm P, L) }{ \partial E} + \frac{\partial [1/\cK_{\df,3}^{\sf iso}(E^*) ] } { \partial E} \bigg )^{\!\! -1} \bigg \vert_{E = E_n^{A_1}(\bm P, L)} \,.
\label{eq:Rnisodef}
\end{equation}
Here all derivatives are evaluated at the energy $E^{A_1}_n(\boldsymbol P, L)$, a solution to the isotropic quantization condition. The quantity $r_n^{\sf iso}$ is real in general, and positive for a physical solution. Thus we can read off the vector $v(E_n,\bm P, \Lambda \mu=A_1, L)$ defined in eq.~(\ref{eq:residueres}),
\begin{equation}
(v_n^{\sf iso})^\dagger = (r_n^{\sf iso})^{1/2} \bra1 F_3\,.
\end{equation}
Here we have chosen the overall phase according to the convention discussed above, so that $v_n^{\sf iso}$ is real. Using eq.~(\ref{eq:MEfinal2}) we now obtain
\begin{align}
\sqrt{2 E_K(\bm P) } L^3 \bra{E_n,\bm P,A_1,L} \cH_W(0) \ket{K, \bm P, L} &=
(r_n^{\sf iso})^{1/2} F^{\sf iso}_3 A^{\sf iso} \,.
\end{align}
This can be massaged into a simple form for determining $A^{\sf iso}$
\begin{multline}
\label{eq:LLsquaredform}
A^{\sf iso}(E^*_n)^2 = 2 E_K(\bm P) L^6 \bra{E_n,\bm P,A_1,L} \cH_W(0) \ket{K, \bm P, L}^2 \\
\times \left( \frac{\partial F_3^{\sf iso} (E, \bm P, L)^{-1} }{ \partial E} + \frac{\partial \cK_{\df,3}^{\sf iso}(E^*)}{\partial E}\right)_{E = E_n^{A_1}(\bm P, L)} \,.
\end{multline}
Thus, in the isotropic approximation, we need to measure the matrix element to only a single three-pion state in order to determine $A^{\sf iso}$ at that energy. In figure \ref{fig:LLPlot} we plot the conversion factor appearing on the second line of this equation for the case of constant $\mathcal K^{\sf iso}_{\text{df},3}$, implying $ \partial \cK_{\df,3}^{\sf iso}(E^*) / \partial E = 0$.

The relationship of $A^{\sf iso}$ to $T_{K3\pi}$ is also substantially simplified in the isotropic approximation. We first note that eq.~(\ref{eq:Tufinal}) simplifies to
\begin{equation}
T^{(u),\sf iso}_{K3\pi,L} = \cL^{(u)}_L\ket1 \frac1{1 + \cK_{\df,3}^{\sf iso} F_3^{\sf iso} } A^{\sf iso}
\,.
\end{equation}
Taking the infinite volume limit as before, we obtain
\begin{equation}
T^{(u), \sf iso}_{K3\pi}(\bm k, \widehat{\bm a}^*) = \cS\left\{ T^{(u),\sf iso}_{K3\pi}(\bm k) \right\}\,,
\label{eq:Tu}
\end{equation}
where
\begin{equation}
T^{(u),\sf iso}_{K3\pi}(\bm k) = \cL^{(u),\sf iso}(\bm k)
\frac{A^{\sf iso}}{1 + \cK_{\df,3}^{\sf iso} F_3^{\infty,\sf iso}}\,.
\label{eq:Tusio}
\end{equation}
Here the momentum dependence arises solely from the final-state interactions in
\begin{equation}
\cL^{(u),\sf iso}(\bm k) = \frac13 - \int_{\bm s} \cD_{23}^{(u,u)}(\bm k,\bm s)
\widetilde \rho_\PV(\bm s)\,,
\end{equation}
where $\cD_{23}^{(u,u)}(\bm k,\bm s)$ still satisfies eq.~(\ref{eq:D23uu}), but now with all quantities restricted to $\ell=m=0$, and
\begin{equation}
F_3^{\infty,\sf iso} = \int_{\bm r} \widetilde \rho_\PV(\bm r) \cL^{(u),\sf iso}(\bm r)
\,.
\end{equation}
In this case, the only integral equation that has to be solved is that for $\cD_{23}^{(u,u)}$, as has been done recently in refs.~\cite{\Hadspecpipipi,\Jackurainteq}. We note that $F_3^{\infty,\sf iso}$ and $\cL^{(u),\sf iso}$ are, in general, complex.

The expressions in the isotropic approximation are sufficiently simple that one can readily combine eqs.~\eqref{eq:LLsquaredform} and \eqref{eq:Tusio} to display the direct relation between the finite-volume matrix element and the physical amplitude. Unpacking the compact notation used above slightly, we reach
\begin{multline}
\vert T^{\sf iso}_{K3\pi}(E^*, m_{12}^2, m_{23}^2) \vert^2 = 2 E_K(\bm P) L^6 \Big|\bra{E_n,\bm P,A_1,L} \cH_W(0) \ket{K, \bm P, L}\Big|^2 \\
\times \bigg \vert \cL^{\sf iso}(E^*, m_{12}^2, m_{23}^2)
\frac{1}{1 + \cK_{\df,3}^{\sf iso}(E^*) F_3^{\infty,\sf iso}(E^*)} \bigg \vert^2 \left( \frac{\partial F_3^{\sf iso} (E, \bm P, L)^{-1} }{ \partial E} + \frac{\partial \cK_{\df,3}^{\sf iso}(E^*)}{\partial E}\right) \,,
\label{eq:LLsquaredME}
\end{multline}
where $E$ (and thus $E^*$) is fixed by the value of finite-volume energy, tuned to $E^* = M_K$ for a physical decay amplitude. We have emphasized that the right-hand side depends on the two squared invariant masses $m_{12}^2$ and $m_{23}^2$, defined by
\begin{align}
m_{12}^2 & = (E - \omega_{ k})^2 - (\boldsymbol P - \boldsymbol k)^2 \,, \\
m_{23}^2 & = (E - \omega_{ a})^2 - (\boldsymbol P - \boldsymbol a)^2 \,,
\end{align}
and have also introduced the symmetrized final-state interaction factor.
\begin{equation}
\cL^{\sf iso}(E^*, m_{12}^2, m_{23}^2) \equiv \cL^{(u),\sf iso}(\bm k) + \cL^{(u),\sf iso}(\bm a) + \cL^{(u),\sf iso}(\bm b) \,.
\end{equation}

At this stage we can comment on the relationship of our result to that of ref.~\cite{Muller:2020wjo}. We expect that the isotropic limit, given in eq.~\eqref{eq:LLsquaredME}, is equivalent to the result of ref.~\cite{Muller:2020wjo}, aside from differences in the schemes used to define the short-distance quantities. Indeed, the equations have the same basic structure, with a contribution resulting from final state interactions (the term involving $\mathcal L^{\sf{iso}}$) and a Lellouch-L\"uscher-like correction factor. Demonstrating the precise equivalence, however, is nontrivial, since our approach based in short-distance quantities, $\Kdf$ and $A_{K3\pi}^\PV$, that are symmetric under particle exchange, whereas the approach of ref.~\cite{Muller:2020wjo} does not symmetrize until the very end. Presumably, the mapping can be determined using the relation between symmetric and asymmetric approaches explained in refs.~\cite{Blanton:2020gha,Blanton:2020jnm}, but this is beyond the scope of the present work.

In closing, we note that eq.~\eqref{eq:LLsquaredME} is analogous to the original Lellouch-L{\"uscher} relation presented in ref.~\cite{\LL}. In particular, the two-particle result is reached by making the replacements
\begin{align}
T^{\sf iso}_{K3\pi}(E^*, m_{12}^2, m_{23}^2) & \longrightarrow T_{K 2\pi}(E) \,, \\
\cL^{\sf iso}(E^*, m_{12}^2, m_{23}^2) & \longrightarrow 1 \,, \\
\cK_{\df,3}^{\sf iso}(E^*) & \longrightarrow \cK_{2}(E) \,, \\
F_3^{\infty,\sf iso}(E^*) & \longrightarrow - i \rho(E) \equiv - i \frac{q}{16 \pi E} \,, \\
F_3^{\sf iso} (E, \bm P, L) & \longrightarrow F(E,L) \,,
\end{align}
where we have also restricted attention to the $\boldsymbol P = \boldsymbol 0$ frame. On the right-hand side we have introduced the physical $K \to \pi \pi$ amplitude $T_{K 2\pi}(E) $, extended to allow for final-state energies different from the kaon mass. We have also used the two-particle K-matrix, $\cK_2$, and the two-particle finite-volume function, $F$, both restricted to the $s$-wave. These are essentially the same quantities as appearing in eq.~\eqref{eq:F3def}, in the definition of $F_3$, but without the implicit sub-threshold regulator used there and without the spectator-momentum index. We have also introduced the two-particle phase-space, $\rho(E)$, with $q = \sqrt{E^2/4 - m^2}$.

Making the indicated substitutions into eq.~\eqref{eq:LLsquaredME} yields
\begin{multline}
\vert T_{K 2\pi}(E) \vert^2 = 2 M_K L^6 \, \Big|\bra{E_n,A_1,L} \cH_W(0) \ket{K, L} \Big|^2 \\
\times \bigg \vert
\frac{1}{1 - i \cK_{2}(E) \rho(E)} \bigg \vert^2 \left( \frac{\partial F (E, L)^{-1} }{ \partial E} + \frac{\partial \cK_{2}(E)}{\partial E}\right) \,.
\end{multline}
Substituting the definitions of the scattering phase $\delta (E)$ and the $L$-dependent, so-called pseudophase $\phi(E,L)$
\begin{equation}
\cK_{2}(E) = \frac{ 16 \pi E \tan \delta(E)}{q} \,, \qquad \qquad F (E, L)^{-1} = \frac{ 16 \pi E \tan \phi(E,L)}{q} \,,
\end{equation}
one can easily reach eq.~(4.5) of ref.~\cite{\LL}, after some algebraic manipulations.

This completes our discussion of the formalism in the context of the simplified theory. We now turn to realistic applications of these results.

\section{Applications to physical processes}
\label{sec:processes}

In this section, we describe the generalization of the previous analysis to processes involving three-pion final states in isosymmetric QCD. This allows our results to be applied to several processes of phenomenological interest: {(i)} the electromagnetic transition $\gamma^* \to 3\pi$, which contributes to the hadronic vacuum polarization piece of the muon's magnetic momentum, $(g-2)_\mu$; {(ii)} the isospin-violation strong decay $\eta \to 3\pi$; and {(iii)} the weak decay $K \to 3\pi$, which has both CP-conserving and violating amplitudes.

The generalization presented here requires the generic three-pion quantization condition derived in ref.~\cite{Hansen:2020zhy}. We start this section by recalling some results from that work, and presenting the generalization of the formulae derived above to the three-pion system. We then describe the specific applications to the three processes listed above.

\subsection{General considerations}
\label{sec:physicalK3pi}

In the derivation in section~\ref{sec:derivation}, the ``kaon'' and ``pion'' fields were taken to be real scalars with separate $\mathbb Z_2$ symmetries. Here we consider the physical kaon and pion fields. The former, which can be either charged or neutral, are complex fields with strangeness conservation playing the role of the $\mathbb Z_2$ symmetry. The pions are represented by a triplet of fields, with two complex fields in the definite charge basis ($\pi^+$ and $\pi^-$) and one real filed ($\pi^0$), with the $\mathbb Z_2$ symmetry being G parity. Both kaons and pions are stable particles in QCD, with masses satisfying the required inequality, eq.~(\ref{eq:range}). The form of the weak Hamiltonian depends on the decay being considered, but its essential property, unchanged from above, is that it annihilates one of the kaons and creates three pions. The new feature is the presence of multiple three-pion intermediate states, e.g. $\pi^+\pi^0\pi^-$ and $\pi^0\pi^0\pi^0$ in the neutral sector, and it is this feature that the derivation of ref.~\cite{\isospin} takes into account.

We stress again that, since the weak interactions are added by hand as external operators, we can choose to separately consider operators that create three and two pions, with G parity ensuring that these two sectors do not mix. We can also consider one at a time operators that create three pions in states of definite isospin. Indeed, the quantization condition of ref.~\cite{\isospin} decomposes into separate results for each choice of total isospin. Finally, we note that, although we couch the discussion in this subsection in terms of the $K \to3\pi$ decay, the essential aspects of the discussion apply equally well if the kaon is replaced by a $\gamma^*$ or $\eta$, and the weak operator is replaced by the electromagnetic current or the isospin-breaking Hamiltonian, respectively.

A generic three-pion state can have total isospin $I=0,1,2$ and $3$. It is, however, important to note that the isospin of any pair of particles is not conserved---for a given total isospin there can be several two-pion subchannels with pairwise interactions. As discussed in ref.~\cite{Hansen:2020zhy}, the following subchannels contribute
\begin{align}
\begin{split}
I = 0\!:&\ \big\{ \ket{\rho \pi}_0 \big\}, \\
I = 1\!:&\ \big\{ \ket{\sigma \pi}_1,\ket{\rho \pi}_1 ,\ket{(\pi\pi)_2 \pi}_1 \big\}, \\
I = 2\!:&\ \big\{ \ket{\rho \pi}_2,\ket{(\pi\pi)_2 \pi}_2 \big\}, \\
I = 3\!:&\ \big\{ \ket{(\pi\pi)_2 \pi}_3 \big\},
\end{split} \label{eq:states}
\end{align}
where ``$\sigma$'',``$\rho$'',``$(\pi\pi)_2$'' label a two-pion combination with isospin 0,1, and 2, respectively, and the subscripts on the kets denotes the total isospin. Explicit expressions for these states for the charge zero ($I_3=0$) sector are given in appendix C of ref.~\cite{Hansen:2020zhy}.

The order of pion fields in each state of eq.~\eqref{eq:states} is a shorthand for the interplay of momentum and isospin assignment. In particular, if we consider asymptotic states with fixed total energy and momentum $(E, \boldsymbol P)$ then the remaining degrees of freedom, $\ell m$ and $\boldsymbol k$, are assigned to the leading pion pair and the third pion field, respectively. As emphasized in section~\ref{sec:recap}, the asymmetric description is natural from the perspective of the finite-volume formalism, since many of the quantities appearing there, in particular $F$, $G$, $\cK_2$ and $F_3$, single out a pion pair in their definition. The result is that there are additional flavor spaces with dimensions one, three, two and one, for $I=0,1,2,3$ respectively. Aside from this feature, and a minor change in notation (to be described below), the forms of the final results in ref.~\cite{\isospin} are the same as those for identical particles reviewed in section~\ref{sec:recap}.

The simplicity of the generalization from three identical particles to three-pion states carries over to the new quantities needed to discuss decay matrix elements. For this reason we only quote the results. We begin with the generalization of the Euclidean correlator $C_{AB,L}(P)$, defined in eq.~(\ref{eq:CABLdef}). The operators $\cA$ and $\cB$ now respectively destroy and create a three-pion state of definite isospin. The expression for this correlator, previously given by eq.~(\ref{eq:CABLres}), now becomes
\begin{equation}
C^{[I]}_{AB,L} = C^{[I]}_{AB,\infty}
- i \mathbf A^{\PV,[I]} \frac{1}{[\mathbf{F}^{[I]}_3]^{-1} - \mathbf{K}^{[I]}_{\text{df},3}}
\mathbf B^{\PV,[I]}\,.
\label{eq:Cisospin}
\end{equation}
The notation for bold-faced quantities is taken over from ref.~\cite{\isospin}: they contain a factor of $i$ compared to those used for identical particles [which explains differences in signs and factors of $i$ compared to eq.~(\ref{eq:CABLres})] and also have an additional index corresponding to the flavor space described above. For example, for $I=1$, the endcap $\mathbf A^{\PV,[I]}$ is a three-dimensional row vector in these indices (in addition to being a row vector in the $k\ell m$ indices), while $\mathbf F_3^{[I]}$ and $\mathbf{K}_{\df,3}^{[I]}$ are $3\times 3$ flavor matrices (as well as being matrices in the $k\ell m$ indices).\footnote{%
One difference compared to ref.~\cite{\isospin} is that the endcaps in that work are matrices in flavor space, while those here are row or column vectors. This reflects the fact that creation and annihilation operators in ref.~\cite{\isospin} were chosen to create three-pion states of all isospins, whereas here we consider single operators with definite three-pion isospin.
}
The explicit expressions for the flavor structure of $\mathbf F_3^{[I]}$ are given in Table 1 of ref.~\cite{\isospin} and we do not repeat them here.

With eq.~(\ref{eq:Cisospin}) in hand, the derivation in section~\ref{subsec:residual} goes over almost verbatim. One uses the same three correlators, eqs.~(\ref{eq:CK})-(\ref{eq:CK3pi}), except for the above-described changes to the kaon field and the three-pion operators. The final result is a generalization of eq.~(\ref{eq:MEfinal2}):
\begin{equation}
\sqrt{2 E_K(\boldsymbol P)}L^{3} \langle E_n^{\Lambda,[I]},\bm P, I,I_3,\Lambda\mu,L |\cH_W(0)| K,\bm P,L \rangle =
{ \mathbf{v} }^\dagger \mathbf{A}_{K3\pi}^{\PV,[I]}
\,.
\label{eq:MEfinaliso}
\end{equation}
The matrix element on the left-hand side is obtained from the lattice simulation with the kaon state having the desired quantum numbers, and $E_n^{\Lambda,[I]}$ being the energy of a three-pion state of chosen isospin and hypercubic-group irrep. We assume that the weak Hamiltonian couples the kaon to this state, for otherwise the equation is trivially satisfied as both sides vanish. On the right-hand side the column vector $\mathbf{v}$ is an abbreviation for
\begin{equation}
\mathbf{v}(E_n^{\Lambda,[I]},\bm P,I,I_3,\Lambda\mu,L)\,,
\end{equation}
which is a row vector having both $\{k\ell m\}$ and flavor indices, and includes a factor of $i$ relative to the $v$ of section~\ref{subsec:residual} in order to cancel the factor of $i$ in $\mathbf{A}_{K3\pi}^{\PV,[I]}$. It is an eigenvector of $[\mathbf{F}_3^{[I]}]^{-1} - \mathbf{K}_{\df,3}^{[I]}$ with vanishing eigenvalue, and is defined by the generalization of eq.~(\ref{eq:residueres}):
\begin{align}
\mathbf R_{\Lambda\mu}^{[I,I_3]}(E_n^{\Lambda,[I]}, \bm P, L)
&= \lim_{P_4 \to iE_n^{\Lambda,[I]}} -(E_n^{\Lambda,[I]} + i P_4)
\mathbb P_{\Lambda\mu}^{[I,I_3]}
\frac{(-i)}{1/\mathbf{F}^{[I]}_3 - \mathbf{K}^{[I]}_{\text{df},3}}
\mathbb P_{\Lambda\mu}^{[I,I_3]}
\equiv \mathbf v\, {\mathbf v}^\dagger\,.
\label{eq:resiso}
\end{align}
We stress that we do {\em not} include a relative factor of $i$ between the definitions of $\mathbf R_{\Lambda\mu}^{[I,I_3]}$ and $\cR_{\Lambda\mu}$ of section~\ref{subsec:residual}. The bold quantity defined here thus differs from the $\cR_{\Lambda\mu}$ only by the addition the flavor index.

The workflow for using eq.~(\ref{eq:MEfinaliso}) is as follows: First, one chooses the initial kaon quantum numbers and the form of $\cH_W$ based on the physical process under consideration. This determines the allowed values of $I$ and $I_3$ for the three-pion final states. Second, one calculates the three-pion energy spectrum for one of the allowed values of $\{I,I_3\}$, using a range of choices of $\bm P$, and picking irreps/rows $\Lambda\mu$ such that the desired $K\to3\pi$ matrix element is nonvanishing. Third, one compares this spectrum to the result from the quantization condition of ref.~\cite{\isospin},
\begin{equation}
\det \! \Big ( [{\mathbf{F}^{[I]}_3]^{-1} - \mathbf{K}^{[I]}_{\text{df},3}} \Big) = 0\,,
\end{equation}
and uses this to determine (a parameterized form of) $\mathbf{K}^{[I]}_{\text{df},3}$. Fourth, with this form in hand one uses eq.~(\ref{eq:resiso}) to determine the vectors $\mathbf v$ for levels that have their energies matched to $E_K(\bm P)$. Finally, one uses eq.~(\ref{eq:MEfinaliso}) to provide a constraint on the row vector $\mathbf A_{K3\pi}^{\PV,[I]}$. By combining several such constraints can determine a (parametrized form of) $\mathbf A_{K3\pi}^{\PV,[I]}$.

The second step---connecting to the physical decay amplitude---also mirrors that for identical particles, which was described in section~\ref{subsec:PVtodecay}. One first introduces an asymmetric finite-volume amplitude that generalizes eq.~(\ref{eq:Tufinal}),
\begin{align}
\mathbf{T}^{[I](u)}_{K3\pi,L}
&= \left(\mathbf F^{[I]}\right)^{-1} \mathbf F^{[I]}_3
\frac{1}{1 - \mathbf K^{[I]}_{\text{df},3} \mathbf F^{[I]}_3 }
\mathbf A_{K3\pi}^{\PV,[I]} \,,
\label{eq:ABiso}
\end{align}
where $\mathbf F^{[I]}$ is $i F/(2\omega L^3)$ tensored with the identity in the corresponding flavor space~\cite{\isospin}. Here again the boldfaced quantity $\mathbf T^{[I](u)}_{K3\pi,L}$ differs from the $T^{(u)}_{K3\pi,L}$ used in section~\ref{subsec:PVtodecay} both by the addition of flavor indices and by a factor of $i$. The physical amplitude is then obtained by taking the appropriate ordered limit and symmetrizing,
\begin{equation}
\mathbf{T}^{[I]}_{K3\pi} =
\mathcal{S} \left\{ \lim_{\epsilon \to 0^+} \lim_{L \to \infty}
\mathbf{T}^{[I](u)}_{K3\pi,L} \right\}\,.
\label{eq:ABiso2}
\end{equation}
This limit leads to integral equations that are simple generalizations of those presented in section~\ref{subsec:PVtodecay}, and which we do not display explicitly. The only subtlety that is introduced by the flavor indices is the need to generalize the definition of symmetrization, as is explained in section 2.3 of ref.~\cite{\isospin}. We stress that the symmetrization here acts on a column vector with a single index, rather than on a matrix as in ref.~\cite{\isospin}.

The results of these steps are the infinite-volume decay amplitudes in the isospin basis. To convert to a measurable amplitude, e.g.~that for $K^+\to \pi^+\pi^0\pi^0$, one must combine the isospin amplitudes appropriately. The results needed to do this are collected in appendix~\ref{app:3pi}. In this regard there is a further subtlety concerning the amplitudes that have a multi-dimensional flavor space, i.e.~those with $I=1$ and $2$. To explain this point (which is not discussed in ref.~\cite{\isospin}) we focus on the example of $I=1$. The result from eq.~(\ref{eq:ABiso2}) is then {\em three} $K\to [3\pi]_{I=1}$ amplitudes, each expressed as a function of the three pion momenta. The issue is that, when one has the full momentum dependence, these three amplitudes are not independent. In fact, as we explain below, one needs to know only two of the three in order to completely reconstruct the $I=1$ amplitude. Similarly, for the $I=2$ case, only one of the two amplitudes is needed. This redundancy does not, however, lead to any simplification in the solution of the integral equations implicit in eq.~(\ref{eq:ABiso2}).

\subsection{The electromagnetic transition $\gamma^* \to 3\pi$}
\label{subsec:gammato3pi}

The electromagnetic process $\gamma^*\to 3\pi$ is of phenomenological interest as it contributes, via the hadronic vacuum polarization (HVP) and the hadronic light-by-light scattering (HLbL), to the anomalous magnetic moment of the muon~\cite{Hoferichter:2014vra,Hoferichter:2018dmo,Hoferichter:2018kwz,Hoferichter:2019mqg,Hoid:2020xjs}. Our formalism allows one to determine the infinite volume amplitude using a finite volume lattice QCD calculation. In particular, although this is not a decay, the results above are readily adapted---one simply takes advantage of the fact that one can allow the final three-particle state to take on any energy and momentum in the relations given above. This then corresponds to a timelike photon with virtuality $q^2 = E^{\Lambda}_n(L, \boldsymbol P)^2 - \boldsymbol P^2$. The analogous two-particle process, $\gamma^* \to \pi \pi$, and its relation to finite-volume matrix elements is discussed in ref.~\cite{Meyer:2011um}.

The replacement of the kaon with a virtual photon simplifies the required lattice calculation. The composite operator $\mathcal B_{K3\pi}(x)$ in eq.~\eqref{eq:BK3pidef} is replaced by the electromagnetic current $\mathcal J_\nu(x)$, and the kaon correlator is not required. We consider here only the part of this current that involves up and down quarks,
\begin{equation}
\mathcal{J}_\nu = \frac{2}{3} \bar u \gamma_\nu u - \frac{1}{3} \bar d \gamma_\nu d \,,
\end{equation}
as this leads to the dominant contribution to $\gamma^*\to 3\pi$. No tuning of the volume is needed to match a given energy; instead, each finite-volume three pion state with appropriate quantum numbers leads to a result for the desired amplitude with photon virtuality given by the energy of the state.

The electromagnetic current contains both isoscalar and isovector parts. The latter has positive G parity and thus, in isosymmetric QCD, couples only to even numbers of pions, and in particular to the $\rho$ resonance. What is of interest here is the isoscalar part,
\begin{equation}
\mathcal J_\nu^0 = \frac16\left( \bar u \gamma_\nu u + \bar d \gamma_\nu d\right)\,,
\label{eq:Jmu0}
\end{equation}
which has negative G parity and thus couples to three pions. The dominant contribution in the energy range of interest for muonic $g-2$ is from the $\omega(782)$ resonance.

The desired amplitude is obtained using the two-step process explained above. Each matrix element obtained from a lattice calculation is related to the intermediate PV amplitude by
\begin{equation}
L^{3/2} \langle E_n^{\Lambda,[0]},\bm P, I=0,\Lambda\mu,L |\mathcal J^0_\nu(0)| 0 \rangle =
{ \mathbf{v} }^\dagger \mathbf{A}_{\gamma3\pi,\nu}^{\PV,[0]}
\,,
\label{eq:MEgamma3pi}
\end{equation}
where $\mathbf v = \mathbf v^{[0]}(E_n^{\Lambda,[0]}, \bm P,I=0,\Lambda\mu,L)$ is obtained from the spectrum of $I=0$ three pion states using eq.~(\ref{eq:resiso}). The irreps $\Lambda$ and rows $\mu$ that lead to nonzero matrix elements depend on the total momentum $\bm P$ and the Lorentz index $\nu$. Note that for $I=0$ the flavor space is one dimensional, so $\mathbf{A}_{\gamma3\pi,\nu}^{\PV,[0]}$ and $\mathbf v$ can be viewed as vectors in $\{k\ell m\}$ space alone. We also comment that the left-hand side of eq.~\eqref{eq:MEgamma3pi} differs from the corresponding results for kaon decays, eqs.~(\ref{eq:MEfinal2}) and (\ref{eq:MEfinaliso}), by the absence of a factor of $(2 E_K(\boldsymbol P) L^3)^{1/2}$. This is because, in contrast to the unit normalized finite-volume kaon state, there is no need to correct the normalization of the vacuum, which matches between the finite- and infinite-volume theories.

To implement eq.~(\ref{eq:MEgamma3pi}), the infinite-volume PV amplitude $\mathbf{A}_{\gamma3\pi,\nu}^{\PV,[0]}$ must be parametrized. This is most easily done by using eq.~(\ref{eq:decompose}) to convert from $\{k\ell m\}$ space to a function of three on-shell momenta, $p_1$, $p_2$ and $p_3$. Up to the overall factor of $i$, the amplitude is a real, smooth function of momenta, antisymmetric under the interchange of any pair of momenta, and transforming as an axial vector.\footnote{%
If the intrinsic negative parity of the pions is included the amplitude transforms as a vector, as required to couple to the electromagnetic current.}
Expanding about threshold as in section~\ref{subsec:threshold} (with $m_K^2 \to q^2$), the general form satisfying these properties is
\begin{equation}
\mathbf{A}_{\gamma3\pi,\mu}^{\PV,[0]}= i \epsilon_{\mu \nu \rho \sigma} p_1^\nu p_2^\rho p_3^\sigma
\left(A_{\gamma3\pi}^{(0)} + A_{\gamma3\pi}^{(2)} \sum_i \Delta_i^2 + \dots \right)
\,.
\label{eq:Agamm3pi}
\end{equation}
Here the $\Delta_i$ are the threshold expansion parameters defined in eq.~\eqref{eq:DeltaiDef}, and the coefficients $A_{\gamma3\pi}^{(n)}$ are functions of $\Delta=q^2/(9 m_\pi^2) -1$. For a consistent threshold expansion, $A_{\gamma3\pi}^{(0)}$ should be a quadratic function of $\Delta$, while $A_{\gamma3\pi}^{(2)}$ should be a constant. The ellipsis represents higher order terms. We observe that the threshold expansion begins at higher order than for the symmetric amplitude discussed in section~\ref{subsec:threshold}.

The second step is to solve the integral equations encoded in the $I=0$ versions of eqs.~(\ref{eq:ABiso}) and (\ref{eq:ABiso2}), which convert $\mathbf{A}_{\gamma3\pi,\nu}^{\PV,[0]}$ into the $\gamma^*\to [3\pi]_{I=0}$ amplitude, $\mathbf T^{[0]}_{\gamma3\pi,\nu}( p_1, p_2, p_3)$. Recalling from ref.~\cite{\isospin} that the $I=0$ state is given by
\begin{equation}
\frac1{\sqrt6} \left(
\ket{\pi^+ \pi^0 \pi^-} - \ket{\pi^0 \pi^+ \pi^-}
+ \ket{\pi^0 \pi^- \pi^+} - \ket{\pi^- \pi^0 \pi^+}
+ \ket{\pi^- \pi^+ \pi^0} - \ket{\pi^+\pi^- \pi^0} \right)\,,
\end{equation}
with the three pions in each ket having the momenta $p_1$, $p_2$ and $p_3$, respectively, and noting that only the $I=0$ amplitude is nonzero, we obtain the physical amplitude as
\begin{equation}
i\cT\left[ \gamma^* \to \pi^+(p_1) \pi^0(p_2) \pi^-(p_3) \right] =
\sqrt{\frac{1}{6}} \mathbf T^{[0]}_{\gamma3\pi,\nu}( p_1, p_2, p_3),
\end{equation}
where the index $\nu$ refers to the polarization of the virtual photon.\footnote{%
This can also be obtained from the bottom row of the matrix $\cR$ given in eq.~(\ref{eq:Rnumerical}). Since only the $I=0$ amplitude is nonzero, the rightmost entry in this row gives the relevant factor.}

\subsection{The isospin-violating strong decay $\eta \to 3\pi$}
\label{subsec:etato3pi}

The decay $\eta \to 3 \pi$ provides an example where our formalism can be used within the context of the strong interactions. The key point is that the $\eta$ is stable in isosymmetric QCD, but can decay to three pions in the presence of isospin violation.\footnote{%
Potential decays to $2\pi$ and $4\pi^0$ that are allowed by G parity and kinematics are forbidden by parity conservation, irrespective of isospin breaking.}
The decay has a very small partial width, $\Gamma(\eta\to 3\pi)\approx 0.7\;$keV~\cite{Zyla:2020zbs}, and can be treated at leading order in an expansion in isospin breaking. Isospin violation in the Standard Model arises both from the up-down quark mass difference in QCD and from electromagnetic effects. Here, however, isospin breaking from QCD dominates, since electromagnetic effects are of second order in $\alpha$ due to the neutrality of the $\eta$. Thus this process is uniquely sensitive to the up-down quark mass difference. We refer the reader to ref.~\cite{Gan:2020aco} for a recent review of the status of phenomenological predictions for these decays.

A natural approach for a first-principles lattice QCD calculation of these decay amplitudes is to simulate isosymmetric QCD with mass term
\begin{equation}
\mathcal H^{\Delta I=0} = \frac{m_u+m_d}2 \left( \bar u u + \bar d d \right)\,,
\end{equation}
but introduce isospin violation through the insertion of the mass difference operator\footnote{%
We note that this method of calculating isospin-violating effects is similar to the perturbative method introduced in refs.~\cite{deDivitiis:2011eh,deDivitiis:2013xla}, but differs in that here we imagine inserting the operator at a single position rather than over the entire volume.}
\begin{equation}
\mathcal{H}^{\Delta I=1} = \frac{m_u - m_d}2 \left( \bar u u - \bar d d \right).
\end{equation}
This brings the calculation into the same class as that for $K\to3\pi$ decays, with the initial kaon replaced by the $\eta$ and $\cH_W$ replaced by $\cH^{\Delta I=1}$. We observe that, although isospin-breaking is being included only at leading order, our formalism includes all rescattering effects due to final state interactions. Thus it provides an alternative to the dispersive methods used in present analyses~\cite{Colangelo:2018jxw,Kampf:2019bkf}.

Since the initial $\eta$ has $I=0$, the final three pion state has $I=1$. Thus to obtain the $\eta\to 3\pi$ amplitude we can use the results of section~\ref{sec:physicalK3pi}, by simply making the replacement $K \to \eta$, and taking $I=1$. In this way, we can use the formalism to determine the intermediate PV amplitude $\mathbf A^{\PV,[1]}_{\eta3\pi}$ and the final, physical amplitude $\mathbf T^{[1]}_{\eta3\pi}$. We note that these amplitudes have a three-dimensional flavor space. For a practical implementation one needs a parametrization of $\mathbf A^{\PV,[1]}_{\eta3\pi}$, and the relation of $\mathbf T^{[1]}_{\eta3\pi}$ to the amplitudes into charged and neutral pions. We provide these results in the remainder of this subsection.

To present the parametrization of $\mathbf A^{\PV,[1]}_{\eta3\pi}$, it is convenient to use a different basis for the flavor space of three-pion states than that of eq.~(\ref{eq:states}). The new basis, which we denote the $\chi$ basis, uses states that lie in irreps of the symmetric group $S_3$ corresponding to permutations of the three particles. It is given by~\cite{\isospin}
\begin{align}
\begin{split}
\big\{ \ket{\chi_s}_1, \ket{\chi_1}_1,& \ket{\chi_2}_1 \big\}
=   \\  &\left\{ \frac{2}{3} \ket{(\pi\pi)_2 \pi}_1 + \frac{\sqrt{5}}{3} \ket{\sigma \pi}_1,\
- \frac{\sqrt{5}}{3} \ket{(\pi\pi)_2 \pi}_1 + \frac{2}{3} \ket{\sigma \pi}_1,\
\ket{\rho \pi}_1
\right\}\,,
\end{split}
\end{align}
where $\ket{\chi_s}$ transforms in the trivial irrep of $S_3$, while $\{\ket{\chi_1},\ket{\chi_2}\}$ transform in the two-dimensional standard irrep. We refer to appendix C in ref.~\cite{Hansen:2020zhy} for explicit expressions for the isospin states, as well as further discussion of the group properties.

We now adapt the results obtained in ref.~\cite{\isospin} for the parametrizations of scattering amplitudes to that of the intermediate PV amplitude. Working to quadratic order in the threshold expansion, we find
\begin{multline}
\mathbf{A}_{\eta3\pi}^{\PV,[1]} =
i \left( A^{ {\sf s},0}_{\eta3\pi} + A^{ {\sf s},1}_{\eta3\pi} \Delta + A^{ {\sf s},2}_{\eta3\pi} \Delta^2
+A^{ {\sf s},2a}_{\eta3\pi} \sum_i \Delta_i^2 \right)
\begin{pmatrix} 1\\ 0 \\0 \end{pmatrix}
\\
+ i \left(A^{ {\sf d},1}_{\eta3\pi} + A^{ {\sf d},2}_{\eta3\pi}\Delta\right)
\begin{pmatrix} 0 \\ P \cdot \xi_1 \\ P \cdot \xi_2 \end{pmatrix}
+ i A^{ {\sf d},2a}_{\eta3\pi}
\begin{pmatrix} 0 \\ (P\cdot \xi_2)^2-(P \cdot \xi_1)^2
\\ 2 P\cdot \xi_1 P \cdot \xi_2 \end{pmatrix}
+ \dots,
\label{eq:APVeta}
\end{multline}
where $A_{\eta3\pi}^{ {\sf s},0}$, etc.~are real coefficients. The notation is as in section~\ref{subsec:threshold}, except for the replacement $m_K\to m_\eta$, and the use of the new quantities
\begin{equation}
\xi_1 = \frac{1}{\sqrt{6}} \left(2p_3 -p_1-p_2 \right) ,
\ \text{ and } \ \xi_2 = \frac{1}{\sqrt{2}} \left(p_2 -p_1 \right) .
\label{eq:xidef}
\end{equation}
The superscripts $\sf s$ and $\sf d$ refer to the ``singlet'' symmetric and ``doublet'' standard irrep of $S_3$, respectively. We observe that the symmetric part of the amplitude begins at leading order in the threshold expansion, while that transforming in the doublet enters only at linear order.

Finally we describe the reconstruction of the decay amplitudes into final states composed of pions with definite charges, which are
\begin{align}
\cT_\eta^{000}(p_1,p_2,p_3) &\equiv \cT[\eta \to \pi^0(p_1) \pi^0(p_2) \pi^0(p_3)]\,,\\
\cT_\eta^{+0-}(p_1,p_2,p_3) &\equiv \cT[\eta \to \pi^+(p_1) \pi^0(p_2) \pi^-(p_3)]\,.
\end{align}
Our formalism yields the $I=1$ amplitude, which, expressed in the $\chi$ basis, is
\begin{equation}
\mathbf{T}^{[1]}_{\eta3\pi}( p_1, p_2, p_3) = i \begin{pmatrix}
\mathcal{T}^{[1]}_{\sf s}( p_1, p_2, p_3) \\ \mathcal{T}_{{\sf d},1}^{[1]}( p_1, p_2, p_3) \\ \mathcal{T}_{{\sf d},2}^{[1]}( p_1, p_2, p_3)
\end{pmatrix}\,.
\label{eq:etaamplitudes}
\end{equation}
The relation between the $\chi$ basis and that involving particles of definite charge is given in eq.~(\ref{eq:Rnumerical}). Using this result, and the fact that the amplitudes for $I=0$, $2$, and $3$ vanish, we obtain
\begin{align}
\cT_\eta^{000}( p_1, p_2, p_3) &= - \sqrt{\frac35} \mathcal{T}_{\sf s}^{[1]}( p_1, p_2, p_3)\,,
\\
\cT_\eta^{+0-}( p_1, p_2, p_3) &= \frac{1}{\sqrt{15}} \mathcal{T}_{\sf s}^{[1]}( p_1, p_2, p_3)
- \frac1{\sqrt{12}} \mathcal{T}_{{\sf d},1}^{[1]}( p_1, p_2, p_3)
+ \frac12 \mathcal{T}_{{\sf d},2}^{[1]}( p_1, p_2, p_3)\,.
\label{eq:etarel2}
\end{align}
We note that all three $I=1$ amplitudes are invariant under the interchange $p_1\leftrightarrow p_3$, so that $\cT^{+0-}_\eta(p_1,p_2,p_3)=\cT^{+0-}_\eta(p_3,p_2,p_1)$, which is consistent with the positive charge conjugation parity of the pseudoscalar mesons.

As noted earlier, the two doublet amplitudes are not independent when one uses the freedom to permute the momenta. A convenient form of this relationship is
\begin{equation}
\mathcal{T}_{{\sf d},2}^{[1]}( p_1, p_2, p_3) = \frac1{\sqrt3} \mathcal{T}_{{\sf d},1}^{[1]}( p_1, p_2, p_3)
+ \frac2{\sqrt3} \mathcal{T}_{{\sf d},1}^{[1]}( p_1, p_3, p_2)\,,
\label{eq:redundancy}
\end{equation}
where we stress that the order of the momentum arguments differs in the last term. Using this result, eq.~(\ref{eq:etarel2}) can be rewritten as
\begin{equation}
\cT_\eta^{+0-}( p_1, p_2, p_3) = \frac{1}{\sqrt{15}} \mathcal{T}_{\sf s}^{[1]}( p_1, p_2, p_3)
+ \frac1{\sqrt3} \mathcal{T}_{{\sf d},1}^{[1]}( p_1, p_3, p_2)\,.
\label{eq:etarel3}
\end{equation}

\subsection{The weak decay $K \to 3\pi$}
\label{subsec:kto3pi}

Finally, we turn to the $K\to3\pi$ decays that are the primary motivation for this work. We have left these processes to the end as they are the most complicated to analyze. The main reason for developing the formalism for a lattice calculation of the $K\to3\pi$ amplitudes is to provide a method for determining the CP-violating contribution, so as to allow further tests of the Standard Model. This is analogous to the situation with $K\to2\pi$ decays, where the well-measured CP-violating quantity $\epsilon'/\epsilon$ can now be predicted reliably in the Standard Model using lattice QCD \cite{Bai:2015nea,Blum:2015ywa,Abbott:2020hxn}.

In the three-particle case, the decay amplitudes are
\begin{align}
\begin{split}
\cT_K^{+00}(p_1,p_2,p_3) &\equiv \cT[K^+ \to \pi^+(p_1) \pi^0(p_2) \pi^0(p_3)]\,,
\\
\cT_K^{-++}(p_1,p_2,p_3) &\equiv \cT[K^+ \to \pi^-(p_1) \pi^+(p_2) \pi^+(p_3)]\,,
\end{split}
\end{align}
together with their charge conjugates, and the neutral kaon amplitudes
\begin{align}
\begin{split}
\cT_{K_S}^{+-0}(p_1,p_2,p_3) &\equiv \cT[K_S \to \pi^+(p_1) \pi^0(p_2) \pi^-(p_3)]\,,
\\
\cT_{K_S}^{000}(p_1,p_2,p_3) &\equiv \cT[K_S \to \pi^0(p_1) \pi^0(p_2) \pi^0(p_3)]\,,
\\
\cT_{K_L}^{+-0}(p_1,p_2,p_3) &\equiv \cT[K_L \to \pi^+(p_1) \pi^0(p_2) \pi^-(p_3)]\,,
\\
\cT_{K_L}^{000}(p_1,p_2,p_3) &\equiv \cT[K_L \to \pi^0(p_1) \pi^0(p_2) \pi^0(p_3)]\,.
\end{split}
\end{align}
In the absence of CP violation, all are nonzero except for $\cT_{K_S}^{000}$. All have been measured except for those for neutral kaon decays to $3\pi^0$ \cite{Zyla:2020zbs}. The effects of CP violation that are measurable at present involve the charged kaon decays. Specifically, CP violation shows up as a difference between Dalitz plot slope parameters in $K^+$ and $K^-$ decays (see ref.~\cite{Cirigliano:2011ny} for a review). Experimentally, these differences are on the edge of observability~\cite{Batley:2007aa,Batley:2010fj}. Phenomenological predictions for CP violating observables achieve a comparatively higher accuracy~\cite{Gamiz:2003pi,Prades:2007ud}. In light of this situation, we focus here on the formalism for the decays of charged kaons, and specifically on the $K^+$ decay. The generalization to the $K^-$ decay is straightforward, and that for the neutral kaon decays is summarized in appendix~\ref{app:K0}.

The operators needed for a lattice study of this process are those of the effective electroweak Hamiltonian, $\cH_W$. The set of operators that are relevant after running to scales below the charm mass is given for instance in refs.~\cite{Buchalla:1995vs,Blum:2001xb}. Since $\cH_W$ contains only operators that change isospin by $1/2$ or $3/2$, the allowed total isospin of the $3\pi$ state is $I=0$, $1$ and $2$. For charged kaons only decays to $I=1$ and $2$ amplitudes are allowed. Using the formalism described above, a lattice calculation can determine (constraints on) the intermediate amplitudes $\mathbf A_{K3\pi}^{\PV,[1]}$ and $\mathbf A_{K3\pi}^{\PV,[2]}$. We stress that this can be done separately for each choice of total isospin, and for the CP-conserving and CP-violating parts of each operator contained in $\cH_W$. To carry this out in practice one needs, as usual, parametrizations of the PV amplitudes. That for $\mathbf A_{K3\pi}^{\PV,[1]}$ is identical in form to the result given for the $\eta\to3\pi$ amplitude in eq.~(\ref{eq:APVeta}), with only the labels on the coefficients changing:
\begin{multline}
\mathbf{A}_{K3\pi}^{\PV,[1]} =
i \left( A^{[1],{\sf s},0}_{K3\pi} + A^{[1],{\sf s},1}_{K3\pi} \Delta + A^{[1],{\sf s},2}_{K3\pi} \Delta^2
+A^{[1],{\sf s},2a}_{K3\pi} \sum_i \Delta_i^2 \right)
\begin{pmatrix} 1\\ 0 \\0 \end{pmatrix}
\\
+ i \left(A^{[1],{\sf d},1}_{K3\pi} + A^{[1],{\sf d},2}_{K3\pi}\Delta\right)
\begin{pmatrix} 0 \\ P \cdot \xi_1 \\ P \cdot \xi_2 \end{pmatrix}
+ i A^{[1], {\sf d},2a}_{K3\pi}
\begin{pmatrix} 0 \\ (P\cdot \xi_2)^2-(P \cdot \xi_1)^2
\\ 2 P\cdot \xi_1 P \cdot \xi_2 \end{pmatrix}
+ \dots,
\label{eq:APVK1}
\end{multline}
The corresponding result for the $I=2$ case is
\begin{equation}
\mathbf{A}_{K3\pi}^{\PV,[2]} = i \left( A^{[2],{\sf d},1}_{K3\pi}
+ A^{[2],{\sf d},2}_{K3\pi} \Delta \right)
\begin{pmatrix}
P \cdot \xi_1 \\ P \cdot \xi_2
\end{pmatrix}
+ i A^{[2],{\sf d},2a}_{K3\pi}
\begin{pmatrix} (P\cdot \xi_2)^2-(P \cdot \xi_1)^2
\\ 2 P\cdot \xi_1 P \cdot \xi_2 \end{pmatrix}
+ \dots
\label{eq:APVK2}
\end{equation}
Here we are using the basis~\cite{\isospin}
\begin{equation}
\left\{ \ket{\chi_1}_2, \ket{\chi_2}_2 \right\}
=\\ \left\{ \ket{(\pi\pi)_2 \pi}_2, \ket{\rho \pi}_2 \right\}\,,
\end{equation}
which is further discussed in appendix~\ref{app:3pi}. We have worked to quadratic order in the expansions of $\mathbf A_{K3\pi}^{[I]}$, since fits to experimentally measured Dalitz plots usually work only to this order.

Given a determination of $\mathbf A_{K3\pi}^{\PV,[1]}$ and $\mathbf A_{K3\pi}^{\PV,[2]}$, the second step of solving the integral equations leads to the decay amplitudes in the isospin basis. There are five amplitudes\footnote{%
There is a potential confusion with the amplitudes for $\eta$ decay that have the same names---see eq.~(\ref{eq:etaamplitudes}). It should, however, be clear from the context to which process the amplitudes apply.}
\begin{equation}
\mathbf{T}^{[1]}_{K3\pi}( p_1, p_2, p_3) = i \begin{pmatrix}
\mathcal{T}^{[1]}_{\sf s}( p_1, p_2, p_3) \\ \mathcal{T}_{{\sf d},1}^{[1]}( p_1, p_2, p_3) \\ \mathcal{T}_{{\sf d},2}^{[1]}( p_1, p_2, p_3)
\end{pmatrix}, \ \ \ \ \ \mathbf{T}^{[2]}_{K3\pi}( p_1, p_2, p_3) = i \begin{pmatrix}
\mathcal{T}_{{\sf d},1}^{[2]}( p_1, p_2, p_3) \\ \mathcal{T}_{{\sf d},2}^{[2]}( p_1, p_2, p_3)
\end{pmatrix}\,,
\label{eq:Kamplitudes}
\end{equation}
although, as above, only one from each doublet is independent. The form of this redundancy is exactly as in eq.~(\ref{eq:redundancy}) for both $I=1$ and $2$. The relationship of the isospin-basis states to those with pions of definite charges is given in appendix~\ref{app:3pi}. Using these results, and simplifying using the redundancy equation (\ref{eq:redundancy}), we find
\begin{align}
\begin{split}
\cT^{+00}(p_1,p_2,p_3) &= - \frac{1}{\sqrt{15}} \cT^{[1]}_{\sf s}(p_1,p_2,p_3)
+ \frac{1}{\sqrt3} \left[ \cT^{[1]}_{{\sf d},1}(p_1,p_2,p_3) + \cT^{[2]}_{{\sf d},1}(p_1,p_2,p_3) \right]
\\&\quad + \frac{1}{\sqrt3} \left[ \cT^{[1]}_{{\sf d},1}(p_1,p_3,p_2) + \cT^{[2]}_{{\sf d},1}(p_1,p_3,p_2) \right],
\end{split} \\
\begin{split}
\cT^{-++}(p_1,p_2,p_3) &=\frac{2}{\sqrt{15}} \cT^{[1]}_{\sf s}(p_1,p_2,p_3)
+ \frac{1}{\sqrt3} \left[ \cT^{[1]}_{{\sf d},1}(p_1,p_2,p_3) - \cT^{[2]}_{{\sf d},1}(p_1,p_2,p_3) \right]
\\&\quad + \frac{1}{\sqrt3} \left[ \cT^{[1]}_{{\sf d},1}(p_1,p_3,p_2) - \cT^{[2]}_{{\sf d},1}(p_1,p_3,p_2) \right],
\end{split}
\end{align}
where we have used the vanishing of the $I=3$ amplitude.

\section{Conclusion}
\label{sec:conc}

In this article we have derived the formalism that allows the study of three-particle decay processes using input from lattice QCD calculations. This generalizes the well-established formalism for two-particle decays developed by Lellouch and L{\"u}scher~\cite{Lellouch:2000pv} and its subsequent extensions. Specifically, our formalism applies for decays in which the three particles are degenerate and spinless, although they do not need to be identical. Thus, in particular, the phenomenologically important $K\to3\pi$ decays are now accessible to lattice methods in the isospin-symmetric limit. Our formalism applies not only to $1\to 3$ decay processes, but also $0\to 3$ transitions in the strong interactions, such as that for $\gamma^*\to 3\pi$, which is relevant for lattice calculations of the hadronic vacuum polarization contribution to muonic $g-2$.

We have divided the presentation into two parts. In the first, given in section~\ref{sec:derivation}, we give a detailed derivation in a simplified theoretical set up in which the ``pions'' are identical. This allows us to focus on the essential new features that are introduced when moving from two to three particles. The derivation is carried out by extending the relativistic three-particle finite-volume formalism for identical scalar particles~\cite{Hansen:2014eka,Hansen:2015zga}. Just as in the relation between the finite-volume spectrum and scattering amplitudes, the relation we find between finite-volume decay matrix elements and physical decay amplitudes requires two steps. In the first, finite-volume matrix elements are used to constrain an infinite-volume but scheme-dependent intermediate quantity, $A^\PV_{K3\pi}$. This quantity plays a role that is analogous to that of $\Kdf$ in the scattering formalism of refs.~\cite{\HSQCa,\HSQCb}. The second step in the formalism is to relate $A^\PV_{K3\pi}$ to the physical decay amplitude, and is analogous to the relation between $\Kdf$ and the physical scattering amplitude~\cite{\HSQCb}. This relation is achieved by solving integral equations in infinite-volume that incorporate the effects of two- and three-particle final state interactions (entering through the two-particle K matrix $\cK_2$ and $\Kdf$, respectively) and leads to a decay amplitude satisfying the constraints of unitarity.

Our derivation is independent of the details of the effective theory, aside from the assumption of a $\mathbb Z_2$ symmetry analogous to G parity. It holds for decays of ``kaons'' with masses up to the first inelastic threshold, $m_K < 5 m_\pi$. The approach is relativistic, implying, for one thing, that the intermediate amplitude $A^\PV_{K3\pi}$ is Lorentz invariant. We use this constraint to develop an expansion of $A^\PV_{K3\pi}$ about threshold.

It is instructive to compare the two and three-particle formalisms in more detail. The first step of our formalism is the analog of the multiplication by the LL factor that is required for two-particle decays involving only a single channel. In particular, the vector $v$ that enters the key relation, eq.~(\ref{eq:MEfinal2}), is determined by a combination of scattering amplitudes and kinematic factors, just as the LL factor is in the two-particle case. The main new feature compared to the two-particle analysis is the need for the second step. In the original LL derivation, this step is essentially replaced by the multiplication by the final-state phase required by Watson's theorem. It is the more complicated nature of three-particle final-state interactions that necessitates the solution of integral equations. Another difference from the original LL result is that, in general, each finite-volume three-particle matrix element serves only to constrain $A^\PV_{K3\pi}$, rather than provide a direct determination. This difference is, however, only due to the simplicity of the set-up considered in the original LL work. If one considers a multiple-channel two-particle system, then each lattice matrix element again only provides a constraint on physical decay amplitudes~\cite{Hansen:2012tf,Briceno:2014uqa,Briceno:2015csa}. Conversely, if we consider the simplest approximation for three-particle scattering and decay amplitude, then, as shown in section~\ref{subsec:isotropic}, only a single finite-volume matrix element is required to determine $A^\PV_{K3\pi}$.

In the second part of our presentation, given in section~\ref{sec:processes}, we generalize the formalism so that it applies for decays to a general three-pion state in isosymmetric QCD. This builds upon our recent generalization of the formalism for three-particle scattering to include all three-pion isospin channels~\cite{\isospin}. It allows us to address phenomenologically relevant processes, and we have discussed in detail three applications: the electromagnetic transition $\gamma^* \to 3\pi$, the isospin-violating decay $\eta \to 3\pi$, and the weak decay $K \to 3\pi$. While most of the features of the formalism for identical particles also hold for three-pion decays, the key difference is that all quantities have an additional isospin index. One impact of this change is that the symmetry properties of the generalization of $A^\PV$ differ from those for identical particles, and we have presented explicit expressions in a threshold expansion that should suffice for realistic calculations.

An important difference between the process $\gamma^* \to 3\pi$ on the one hand and the decays $\eta \to 3\pi$ and $K \to 3\pi$ on the other, is that the latter two have a clear physical interpretation only when the initial and final state energies match, whereas the virtual photon transition is meaningful for all final state energies. However, the formalism presented here also holds for matrix elements in which the kinematics are not perfectly matched. In practice, this freedom can be used to extract $A^\PV_{K3\pi}$ as a function of the final state energy, e.g.~by fitting to multiple closely spaced states. This could be useful both for giving stronger constraints on the target amplitude and for interpreting the value, including the role of resonance enhancement in the amplitude, by considering the result for energies away from physical kinematics.

Although a controlled computation of the $K \to 2\pi$ decay amplitude using lattice QCD has only been achieved very recently~\cite{Abbott:2020hxn}, we are hopeful that the extension to $K \to 3\pi$ decays can be undertaken in the next few years. This will require a program of calculations of the finite-volume three-pion spectrum with all allowed total isospins, in addition to the calculation of the finite volume $K\to3\pi$ matrix elements. We note that work on the second step of our formalism---which requires solving integral equations---can begin independently of lattice simulations, since the methods required do not depend on the functional form of the necessary input quantities ($\cK_2$, $\Kdf$ and $A^\PV_{K3\pi}$). Indeed, methods for solving the closely-related integral equations required for three-particle scattering are under active development~\cite{Jackura:2020bsk,Hansen:2020otl}.

Finally, we note that further generalizations of the formalism derived here will be needed to allow lattice calculations of all three-particle decay amplitudes of interest. For example, to address isospin breaking in $K\to3\pi$ decays requires formalism for three nondegenerate particles, as well as for multiple, nondegenerate channels. The recent extension of the three-particle quantization condition to the case of nondegenerate particles is a first step in this direction~\cite{Blanton:2020gmf}.

\acknowledgments
We thank Ra\'ul Briceño and Toni Pich for useful discussions. The work of MTH is supported by UK Research and Innovation Future Leader Fellowship MR/T019956/1. FRL acknowledges the support provided by the European projects H2020-MSCA-ITN-2019//860881-HIDDeN, the Spanish project FPA2017-85985-P, and the Generalitat Valenciana grant PROMETEO/2019/083. The work of FRL also received funding from the European Union Horizon 2020 research and innovation program under the Marie Sk{\l}odowska-Curie grant agreement No. 713673 and ``La Caixa'' Foundation (ID 100010434, LCF/BQ/IN17/11620044). FRL also acknowledges financial support from Generalitat Valenciana through the plan GenT program (CIDEGENT/2019/040). The work of SRS is supported in part by the United States Department of Energy (USDOE) grant No. DE-SC0011637.

\appendix

\section{Proof that $A_3' = A_3^\dagger$\label{app:A3pISA3dag}}

In this appendix, we prove that the quantities $A_3$ and $A_3'$, introduced in eq.~(\ref{eq:CL3res}), are related by hermitian conjugation, provided that the same is true of the two operators entering the corresponding correlation function, eq.~\eqref{eq:CL3}. This result is required to reach eq.~(\ref{eq:C3piHS}), which is used, in turn, to derive the main result of section~\ref{sec:derivation}.

A constructive definition of the quantities $A_3$ and $A_3'$ is provided in ref.~\cite{\HSQCa}, but it is cumbersome and difficult to use in proving basic relations. Therefore, here we find it easier to pursue an indirect method. Our approach is in the spirit of ref.~\cite{\HSQCb} in which $\mathcal K_{\df, 3}$ is related to the physical scattering amplitude via a finite-volume quantity, without making direct use of the complicated constructive definition of ref.~\cite{\HSQCa}.

The key idea is to use the relation between $A_3$, $A_3'$ and their corresponding finite-volume decay amplitudes. To define the latter we first introduce matrix elements defined in terms of physical, asymptotic three-particle states:
\begin{align}
T' (E, \boldsymbol k, \widehat{\boldsymbol a}^\star)&
= \langle0 | \sigma(0) \vert 3 \pi, \text{in} \rangle \,,
\\
T(E, \boldsymbol k, \widehat {\boldsymbol a}^\star) &
= \langle 3 \pi, \text{out} \vert \sigma^\dagger(0)|0 \rangle\,,
\label{eq:T3pidef}
\end{align}
where the arguments on the left-hand side provide a description of the three incoming or outgoing pions, as described in the text following \eqref{eq:symTu}. Starting from these, one can give diagrammatic definitions of $T^{(u)}_L $ and $T^{\prime(u)}_L $, the asymmetric finite-volume decay amplitudes corresponding to $A_3$ and $A_3'$ respectively. For concreteness, we focus on $T^{(u)}_L$; the argument for $T^{\prime(u)}_L $ is analogous. The definition of $T^{(u)}_L$ is essentially the same as that for $T^{(u)}_{K3\pi,L}$ given in section~\ref{subsec:PVtodecay}, except that the initial amputated kaon propagator is absent, so that the initial kaon state in eq.~(\ref{eq:TK3pidef}) is replaced in eq.~(\ref{eq:T3pidef}) with the vacuum. In words, $T^{(u)}_L(E,\boldsymbol k,\widehat{\boldsymbol a}^*)$ is the asymmetric finite-volume vacuum to three pion amplitude in which, if the final interaction involves a $2\to2$ Bethe-Salpeter kernel, then $\bf k$ is the momentum assigned to the spectator, and if the final interaction involves the $3\to3$ kernel, the diagram is multiplied by $1/3$. The amplitude $T(E, \boldsymbol k, \widehat {\boldsymbol a}^\star)$ in eq.~(\ref{eq:T3pidef}) is then obtained by taking the appropriate $L \to \infty$ limit and symmetrizing, just as for $T_{K3\pi}$ in eqs.~\eqref{eq:limTu}-\eqref{eq:symTu} of the main text. 

From the analysis given in section~\ref{subsec:PVtodecay}, it then follows that
\begin{align}
T^{(u)}_L &= X_L A_3\,,
\qquad
X_L = \cL_L^{(u)} \frac1{1+\cK_{\df,3} F_3} \,,
\label{eq:TD}
\end{align}
with $T^{(u)}_L$ a column vector in $\{k\ell m\}$ space, and $\cL_L^{(u)}$ is given in eq.~\eqref{eq:LLu}. This has exactly the same structure as eq.~(\ref{eq:Tufinal}), with $A_{K3\pi}^\PV$ replaced here with $A_3$. A similar analysis leads to
\begin{align}
T'^{(u)}_L &= A_3' X_R \,,
\qquad
X_R = \frac1{1+ F_3 \cK_{\df,3} } \cR_L^{(u)} \,,
\label{eq:TDprime}
\end{align}
with $T'^{(u)}_L$ a row vector in $\{k\ell m\}$ space, and $\cR_L^{(u)}$ given in eq.~\eqref{eq:RLu}. The first key observation is now that
\begin{equation}
X_R = X_L^\dagger\,,
\end{equation}
which follows because $\cL_L^{(u)\dagger} = \cR_L^{(u)}$, $F_3 = F_3^\dagger$ and $\cK_{\df,3} = \cK_{\df,3}^\dagger$. These results themselves follow from the hermiticity of the building blocks $F$, $\cK_2$ and $(2\omega L^3)^{-1}G$.

The second key relation that we need is
\begin{equation}
T'^{(u)}_L = (T^{(u)}_L )^\dagger \,,
\label{eq:TLpISTldag}
\end{equation}
which follows directly from the diagrammatic definitions of $T^{(u)}_L $ and $T'^{(u)}_L $ [without reference to eqs.~\eqref{eq:TD} and \eqref{eq:TDprime}], assuming T and P invariance of the effective field theory, and P invariance of the operator $\sigma$ 
 (ignoring the intrinsic parity of the pion).
To make the argument, we first we note that, aside from phases arising from the operators $\sigma^\dagger$ and $\sigma$, each diagram contributing to $T^{(u)}_L $ and $T'^{(u)}_L $ is real. This is because we are working in finite volume. One way to show this result is to evaluate diagrams using time-ordered perturbation theory, in which case the only source of imaginary contributions is the $i\epsilon$ in the energy denominators. But in finite volume, the sums over spatial momenta do not require that the poles from these denominators be regulated, so that $\epsilon$ can be set to zero. Next we note that T invariance implies the relation $T'^{(u)}_L(E,{\boldsymbol k},\widehat{\boldsymbol a}^\star) =T^{(u)}_L(E,-{\boldsymbol k},-\widehat{\boldsymbol a}^\star)^*$, where complex conjugation is only needed because of possible phases arising from $\sigma^\dagger$ and $\sigma$. Now, using parity invariance, we have that $T^{(u)}_L(E,-{\boldsymbol k},-\widehat{\boldsymbol a}^\star) = T^{(u)}_L(E,{\boldsymbol k},\widehat{\boldsymbol a}^\star)$. Finally, decomposing into the $\{k\ell m\}$ basis, and taking into account that $T'^{(u)}_L$ is a row vector and $T^{(u)}_L$ a column vector, we obtain eq.~(\ref{eq:TLpISTldag}).

Combining eqs.~\eqref{eq:TD}, \eqref{eq:TDprime} and \eqref{eq:TLpISTldag} yields
\begin{equation}
A'_3 X_R = A_3^\dagger X_R \,.
\label{eq:nearlydone}
\end{equation}

The final step is to note that, for any total energy $E$, $X_R$ is well-defined and invertible away from a discrete set of values of $L$ for which one of its eigenvalues vanishes or diverges. 
Away from these ``singular'' values of $L$, we can apply the inverse of $X_R$ to both sides of eq.~(\ref{eq:nearlydone}), and conclude that $A'_3 = A_3^\dagger$.
This demonstrates the desired equality  for all values of 
the spectator momentum $\boldsymbol k$ that lie in the finite-volume sets of the nonsingular
values of $L$.
Assuming that the nonsingular values of $L$ form a dense set,
then, given that $A_3$ and $A'_3$ are continuous functions of the spectator momentum, 
we find that $A'_3=A_3^\dagger$ in general.

\section{Alternative partial derivation following Lellouch-L{\"u}scher method}
\label{app:LLderivation}

Here we follow the approach of ref.~\cite{Lellouch:2000pv}, which provides an alternative to the first step of the derivation,  which is 
 presented in the main text in section~\ref{subsec:residual}. We consider the same theory as in section~\ref{sec:derivation} but now imagine determining the finite-volume spectrum in the two-pion and three-pion sectors in the presence of the weak interaction, with Hamiltonian density $\mathcal H_W(x)$. These sectors are still decoupled in the presence of $\mathcal H_W$, differing by whether the total number of particles is even or odd. The logic of the approach is that the weak interactions shift the spectrum, beginning at linear order, and these shifts can be calculated in two ways: (i) from the finite-volume matrix element; (ii) using the quantization condition, due to a shift in the infinite-volume interactions that depends on the infinite-volume decay amplitude. Comparing the two results for the shift leads to the desired relation.
We stress that throughout this section we drop contributions of quadratic or higher
 order in $\mathcal H_W$ from all equations.

We begin with the two-pion sector. A key distinction here, as compared to the $K \to \pi \pi$ case of ref.~\cite{Lellouch:2000pv}, is that $\mathcal H_W$ only couples the single kaon to states with G parity minus. Thus, the lightest new intermediate state coupling to $\pi \pi$ via the weak interactions is the $K\pi$ state, which, given the constraint eq.~(\ref{eq:range}), has a CMF energy $E_2^*$ that exceeds $4 m_\pi$. It follows that the spectrum in the energy range $ E_2^* < 4 m_\pi$ will only be shifted by second-order weak processes involving off-shell intermediate $K\pi$ states. Since we work at linear order, these can be ignored. Thus the energy levels are unchanged, which, using the two-particle quantization condition, implies that the two-particle scattering amplitude $\cM_2$ is also unchanged. The latter result can also be seen by studying the modifications to this amplitude directly in infinite volume.

The situation is different in the three-pion sector. Here the lightest new intermediate state consists of a single kaon, and this is kinematically allowed; see again (\ref{eq:range}). Levels away from the kaon energy will be shifted only at second order in perturbation theory. However, if the volume is tuned so that there is a three-pion level in the theory without weak interactions whose CMF energy matches that of a finite-volume kaon, then we must use degenerate perturbation theory at leading order.%
\footnote{
The difference between finite- and infinite-volume kaon energies is exponentially suppressed in $L$
and thus neglected in this derivation. 
Therefore, strictly speaking, the approach described in this appendix
is equally valid whether one tunes the three-pion level to the finite- or the infinite-volume kaon energy.
However, in practice, the tuning should be to the finite-volume kaon as this is the quantity
 available in the lattice calculation.
 }
We consider here only a rotationally invariant, local form of $\mathcal H_W(x)$ [such as that of eq.~(\ref{eq:HWsimple})]. In this case, only the trivial irrep of the appropriate little group will be coupled to the kaon and thus only the tuned QCD level in this irrep is relevant. The degenerate sector is thus $(\ket{K,\bm P, L},\ket{E_n,\bm P, A_1, L})$, and the Hamiltonian restricted to this sector is
\begin{equation}
\begin{pmatrix}
E_K(\bm P) & M(\bm P) \\ M^*(\bm P) & E_K(\bm P)
\end{pmatrix}\,,
\quad M(\bm P)= L^3 \bra{E_n,\bm P,A_1,L}\cH_W(0) \ket{K,\bm P, L}\,,
\label{eq:Mdef}
\end{equation}
where the factor of $L^3$ arises due to the difference between Hamiltonian and Hamiltonian density. Diagonalizing, we obtain the energies to first order in $\mathcal H_W$,
\begin{equation}
E_K(\bm P) \to E_K^\pm(\bm P)\equiv E_K(\bm P) \pm |M(\bm P)|\,.
\end{equation}
This is the first result for the energy shifts.

To obtain the second result for the shifts we begin by noting that, when the total CMF energy $E_3^*$ lies within $\cO(\mathcal H_W)$ of $m_K$, the three-particle scattering amplitude is changed at linear order in $\mathcal H_W$. This is because of the nearly on-shell process $3\pi \to K \to 3 \pi$, which leads to
\begin{align}
i \delta \cM_3 ( E_3^* ) & \equiv i \cM^{[\mathcal H_W \neq 0]}_3 ( E_3^* ) - i \cM^{[\mathcal H_W = 0]}_3 ( E_3^* ) \,, \\ & = \bra{3\pi,{\rm out}}\left[ -i\cH_W(0)\right] \ket{K,\bm P}
\frac{i}{E_3^{*2} - m_K^2+i\epsilon}
\bra{K,\bm P} \left[-i \cH_W(0)\right] \ket{3\pi,{\rm in}}
\,.
\end{align}
where we have used the superscripts $[\mathcal H_W \neq 0]$ and $[\mathcal H_W = 0]$ to indicate whether the $3 \pi \to K \to 3 \pi$ transition is present or absent. Here the dependence on the initial and final pion momenta is implicit. Although this appears to be of second order in $\mathcal H_W$, the denominator of the propagator is
\begin{align}
E_3^{*2}-m_K^2 & = E_3(\bm P)^2 - E_K(\bm P)^2 \,, \\
& = 2 E_K(\bm P) \big [ E_3(\bm P) - E_K(\bm P) \big ] + \mathcal O\big [ (E_3(\bm P)-E_K(\bm P))^2 \big ] \,,
\end{align}
and thus of $\cO(\mathcal H_W)$ for $E_3(\bm P)=E_K^\pm(\bm P)$.
It follows that the difference between the perturbed and unperturbed amplitudes at the shifted finite-volume energy is $\cO(\mathcal H_W)$:
\begin{align}
\delta_\pm \cM_3 & \equiv \delta \cM_3 \big ( [E_K^{\pm}(\bm P)^{2} - \bm P^2 ]^{1/2} \big ) \,, \\[5pt] & = \mp \frac{\bra{3\pi,{\rm out}} \cH_W(0) \ket{K,\bm P}
\bra{K,\bm P} \cH_W(0) \ket{3\pi,{\rm in}}}
{2 E_K(\bm P) |M(\bm P)|}
\,.
\label{eq:DeltaM3}
\end{align}

Our next task is to determine the shift in $\cK_{\df,3}$ that corresponds to that in $\cM_3$, for the former is the quantity that enters the quantization condition. For the sake of brevity, we write the following expressions in terms of finite-volume quantities, with the $L\to\infty$ limit implied. We use the expression for $\cM_{3,L}^{(u,u)}$, eq.~(\ref{eq:M3Luu}), but need keep only the second, divergence-free term, since $\cD^{(u,u)}$ does not depend on $\cK_{\df,3}$:
\begin{align}
\delta \cM_3 &= \cS\left\{ \delta \cM_{\df,3,L}^{(u,u)} \right\}\,,
\label{eq:DeltaM3a}
\\
\cM_{\df,3,L}^{(u,u)} &= \cL^{(u)}_L \frac1{1+\cK_{\df,3} F_3} \cK_{\df,3} \cR^{(u)}_L\,,
\\
\delta \cM_{\df,3,L}^{(u,u)} &=
\cL^{(u)}_L \frac1{1+\cK_{\df,3} F_3} \delta \cK_{\df,3} \frac1{1+ F_3 \cK_{\df,3}} \cR^{(u)}_L\,.
\label{eq:DeltaMdf3L}
\end{align}
Next we use eq.~(\ref{eq:Tufinal}) for the decay amplitude, and the conjugate result for the $3\pi\to K$ amplitude, to rewrite eq.~(\ref{eq:DeltaM3}) as
\begin{equation}
\delta_{\pm} \cM_3 = \mp \cS\left\{
\cL_L^{(u)} \frac1{1+\cK_{\df,3} F_3}
\frac{A_{K3\pi}^\PV A_{K3\pi}^{\PV\,\dagger}}{2 E_K |M|}
\frac1{1+F_3\cK_{\df,3}} \cR_L^{(u)} \right\}
\,,
\label{eq:DeltaM3RFT}
\end{equation}
where we have suppressed the $\bm P$ dependence in $E_K$ and $M$. Matching eqs.~(\ref{eq:DeltaM3a}) and (\ref{eq:DeltaMdf3L}) with eq.~(\ref{eq:DeltaM3RFT}), we find
\begin{equation}
\delta_\pm \cK_{\df,3} = \mp \frac{A_{K3\pi}^\PV A_{K3\pi}^{\PV\,\dagger}}{2 E_K |M|}\,.
\label{eq:Kdf3shift}
\end{equation}
The outer product structure reflects the factorization of the residue at the pole in $\cM_3$.

The final step is to enforce the quantization condition with the shifted amplitude at the shifted energies. To this end we define
\begin{equation}
A(E) \equiv F_3(E, \bm P, L)^{-1} + \cK_{\df,3}(E^*) \,.
\end{equation}
Then the unshifted quantization condition can be written as $\det[A(E_K)] = 0$,
and the shifted version as
\begin{equation}
\det [A(E_K^\pm) + \delta_\pm \cK_{\df,3}] = \det [A(E_K) + \delta_\pm A] = 0 \,,
\end{equation}
where we have introduced
\begin{equation}
\label{eq:dAdef}
\delta_{\pm} A = \pm |M| \frac{dA}{dE}\Bigg|_{E_K} + \delta_\pm \cK_{\df,3} \,.
\end{equation}
Recalling that $v$ is the eigenvector of $A(E_K)$ with vanishing eigenvalue, and defining $v + \delta_{\pm} v$ as the corresponding eigenvector for $A(E_K) + \delta_\pm A$, we have
\begin{equation}
(v^\dagger + \delta_{\pm} v^\dagger) \cdot [A(E_K) + \delta_\pm A] \cdot (v + \delta_{\pm} v) = 0 \,. \label{eq:deltaEW}
\end{equation}
Multiplying out this result, using $A(E_K) \cdot v = 0 =v^\dagger \cdot A(E_K)$, and using the fact that the left-hand side of eq.~(\ref{eq:deltaEW}) must vanish order by order in $\cH_W$ (in particular at linear order) yields
\begin{equation}
v^\dagger \cdot \delta_\pm A \cdot v = 0 \,.
\end{equation}

Substituting eqs.~\eqref{eq:Kdf3shift} and \eqref{eq:dAdef} then gives
\begin{equation}
\label{eq:almost}
|M(\bm P)| \bigg [v^\dagger \cdot \frac{dA}{dE}\Bigg|_{E_K} \cdot v \bigg ]= v^\dagger \cdot \frac{A_{K3\pi}^\PV A_{K3\pi}^{\PV\,\dagger}}{2 E_K(\bm P) |M(\bm P)|} \cdot v \,.
\end{equation}
To evaluate the quantity in square brackets we use eqs.~\eqref{eq:residuedef} and \eqref{eq:residueres} of the main text, which imply
\begin{equation}
A(E) = (E - E_K) \frac{v v^\dagger}{\vert v \vert^4 } + X(E) \,,
\end{equation}
where the first term results from
\begin{equation}
A(E)^{-1} = \frac{v v^\dagger}{E - E_K} + \mathcal O \big [ (E - E_K)^0 \big ] \,,
\end{equation}
and $X(E)$ arises from the non-singular part of $A^{-1}$. 
Here we only require that $X(E)$ satisfies $v \cdot X(E) \cdot v^\dagger = 0$. This relies on the fact that the eigenvectors of $A(E_K)$ form a complete set that can be used for any $A(E)$. Then $X(E)$ is built from the sum over all eigenvector pairs $\bm e^{(i)} \bm e^{(j)\dagger}$, weighted by $E$-dependent coefficients, with at least one of the two vectors $\bm e^{(i)}$ and $\bm e^{(j)}$ orthogonal to $v$. From this it immediately follows that
\begin{equation}
\label{eq:one}
v^\dagger \cdot \frac{dA}{dE}\Bigg|_{E_K} \cdot v = 1 \,.
\end{equation}
Finally, inserting eqs.~\eqref{eq:Mdef} and \eqref{eq:one} into eq.~\eqref{eq:almost}, we obtain
\begin{equation}
|v^\dagger A_{K3\pi}^\PV |^2 = 2 E_K(\bm P) L^6
\left|\bra{E_n,\bm P,A_1,L}\cH_W(0) \ket{K,\bm P, L}\right|^2\,.
\end{equation}
This agrees with eq.~(\ref{eq:MEfinal2}) in the main text.

\section{Relations between three-pion states}\label{app:3pi}

In ref.~\cite{Hansen:2020zhy}, we provided the isospin decomposition for all neutral ($I_3=0$) three-pion states, and described the decomposition into irreducible representations of the group $S_3$. Here we provide a result for the neutral sector not given explicitly in ref.~\cite{\isospin}, since this is needed in the discussion of the $\gamma^*\to3\pi$ and $\eta\to 3\pi$ processes. In addition, we generalize the results to the charge 1 ($I_3=1$) sector, as these are needed in the discussion of $K^+$ decays.

The first result is for the matrix $\cR$ defined by
\begin{equation}
\left (
\begin{array}{c}
\ket{-\,0\,+}\\[5pt]
\ket{0-+}\\[5pt]
\ket{-+0}\\[5pt]
\ket{0\ \,0\ \, 0}\\[5pt]
\ket{+-0}\\[5pt]
\ket{0+-}\\[5pt]
\ket{+\, 0\,-}
\end{array}
\right )
= \mathcal R \cdot
\left (
\begin{array}{c}
\ket{(\pi \pi)_2 \pi}_3 \\[5pt]
\ket{(\pi \pi)_2 \pi}_2 = \ket{\chi_1}_2 \\[5pt]
\ket{\rho \pi}_2 =\ket{\chi_2}_2\\[5pt]
\ket{\chi_s}_1 \\[5pt]
\ket{\chi_1}_1 \\[5pt]
\ket{\chi_2}_1 \\[5pt]
\ket{\rho\pi}_0
\end{array}
\right )
\,,
\label{eq:Cdef}
\end{equation}
where we are using the shorthands
\begin{equation}
\ket{-\,0\,+} \equiv \ket{\pi^-(p_1) \pi^0(p_2) \pi^+ (p_3)}\,,\quad
\ket{+\,0\,-} \equiv \ket{\pi^+(p_1) \pi^0(p_2) \pi^- (p_3)}\,,\ \ {\rm etc.}
\end{equation}
We find
\begin{equation}
\mathcal R = \left(
\begin{array}{rrrrrrr}
\frac{1}{\sqrt{10}} & -\frac12 & -\frac1{\sqrt{12}} & \frac1{\sqrt{15}} & - \frac1{\sqrt{12}} & \frac12 & -\frac1{\sqrt6} \\[4pt]
\frac{1}{\sqrt{10}} & -\frac12 & \frac1{\sqrt{12}} & \frac1{\sqrt{15}} & -\frac1{\sqrt{12}} & -\frac12 & \frac1{\sqrt6} \\[4pt]
\frac{1}{\sqrt{10}} & 0 & - \frac2{\sqrt{12}} & \frac1{\sqrt{15}} & \frac2{\sqrt{12}} & 0 & \frac1{\sqrt6} \\[4pt]
\frac{2}{\sqrt{10}} & 0 & 0 &-\frac3{\sqrt{15}} & 0 & 0 & 0 \\[4pt]
\frac{1}{\sqrt{10}} & 0& \frac2{\sqrt{12}} & \frac1{\sqrt{15}} & \frac2{\sqrt{12}} & 0 & -\frac1{\sqrt6} \\[4pt]
\frac{1}{\sqrt{10}} & \frac12 & - \frac1{\sqrt{12}} & \frac1{\sqrt{15}} & -\frac1{\sqrt{12}} & -\frac12 & -\frac1{\sqrt6} \\[4pt]
\frac{1}{\sqrt{10}} & \frac12 & \frac1{\sqrt{12}} & \frac1{\sqrt{15}} & - \frac1{\sqrt{12}} & \frac12 & \frac1{\sqrt6}
\end{array}
\right) \,.
\label{eq:Rnumerical}
\end{equation}
We use the last row of $\cR$ in sections~\ref{subsec:gammato3pi} and
\ref{subsec:etato3pi}.

\medskip
We now turn to the charge 1 sector of three pions, giving our conventions for the states and the relation between the isospin and definite-charge bases. In this sector, the total isospin can only be $I=1$, $2$ or $3$, with degeneracies $3$, $2$, $1$, respectively~\cite{\isospin}. The $S_3$ irreps that appear are the symmetric irrep, labeled $\ket{\chi_s}_I$, and the two-dimensional standard irrep, labeled $\{ \ket{\chi_1}_I,\ket{\chi_2}_I \}$.

The relation to the states in the basis with definite isospin for the first pair is
\begin{align}
\ket{\chi_s}^+_3 &= \ket{(\pi\pi)_2 \pi}_3^+ \\
\ket{\chi_1}^+_2&= \ket{(\pi\pi)_2 \pi}_2^+ \\
\ket{\chi_2}^+_2 &=\ket{\rho \pi}_2^+ , \\
\ket{\chi_s}^+_1 &= \frac{2}{3} \ket{(\pi\pi)_2 \pi}_1^+ + \frac{\sqrt{5}}{3} \ket{\sigma \pi}_1^+ , \\
\ket{\chi_1}^+_1 &=-\frac{\sqrt{5}}{3} \ket{(\pi\pi)_2 \pi}_1^+ + \frac{2}{3} \ket{\sigma \pi}_1^+ , \\
\ket{\chi_2}^+_1 &=\ket{\rho \pi}_1^+
\,.
\end{align}
From this, the relation to the states composed of pions of definite charges is simple to obtain. What we need in section~\ref{subsec:kto3pi} is this inverse of this relation,
\begin{equation}
\left (
\begin{array}{c}
\ket{+\,\,0\,\,0}\\[5pt]
\ket{0\,+\,0}\\[5pt]
\ket{0\,\,\,0\,\,+}\\[5pt]
\ket{-++}\\[5pt]
\ket{+-+}\\[5pt]
\ket{++-}
\end{array}
\right )
= \mathcal R_1 \cdot
\left (
\begin{array}{c}
\ket{\chi_s}_3^+ \\[5pt]
\ket{\chi_1}_2^+ \\[5pt]
\ket{\chi_2}_2^+\\[5pt]
\ket{\chi_s}_1^+ \\[5pt]
\ket{\chi_1}_1^+ \\[5pt]
\ket{\chi_2}_2^+
\end{array}
\right )
\,,
\label{eq:R1def}
\end{equation}
where
\begin{equation}
\mathcal R_1 = \left(
\begin{array}{rrrrrr}
\frac{2}{\sqrt{15}} & \frac1{\sqrt{12}} & \frac12 & -\frac1{\sqrt{15}} & \frac1{\sqrt{12}} & \frac12 \\[4pt]
\frac{2}{\sqrt{15}} & \frac1{\sqrt{12}} & -\frac12 & -\frac1{\sqrt{15}} & \frac1{\sqrt{12}} & -\frac12 \\[4pt]
\frac{2}{\sqrt{15}} & -\frac2{\sqrt{12}} & 0 & -\frac1{\sqrt{15}} & -\frac2{\sqrt{12}} & 0 \\[4pt]
\frac{1}{\sqrt{15}} & -\frac1{\sqrt{12}} & -\frac12 & \frac2{\sqrt{15}} & \frac1{\sqrt{12}} & \frac12 \\[4pt]
\frac{1}{\sqrt{15}} & -\frac1{\sqrt{12}} & \frac12 & \frac2{\sqrt{15}} & \frac1{\sqrt{12}} & -\frac12 \\[4pt]
\frac{1}{\sqrt{15}} & \frac2{\sqrt{12}} & 0 & \frac2{\sqrt{15}} & -\frac2{\sqrt{12}} & 0 \\[4pt]
\end{array}
\right) \,.
\label{eq:R1numerical}
\end{equation}

\section{Formalism for $K^0\to 3\pi$ decays}\label{app:K0}

For completeness, we collect here the results needed to apply the formalism to the decays of neutral kaons. We do so for the $K^0$ decay. That for $\overline K^0$ decay is identical in form, and by forming appropriate combinations one can determine the amplitudes for $K_S$ and $K_L$ decays.

The major change compared to $K^+$ decays is the presence of the $I=0$ final state in addition to those with $I=1$ and $2$. The parametrization of the intermediate PV $I=0$ amplitude requires an antisymmetric combination of the pion momenta that is a Lorentz invariant. In terms of the parameters defined in section~\ref{subsec:threshold}, we find that the leading term is of cubic order in the threshold expansion,
\begin{equation}
\mathbf A_{K3\pi}^{\PV,[0]} = i A_{K3\pi}^a \left[
\Delta_3^2 (\Delta_1-\Delta_2) + \Delta_1^2 (\Delta_2-\Delta_3) + \Delta_2^2 (\Delta_3-\Delta_1)
\right] + \dots \,.
\end{equation}
The parametrizations of the $I=1$ and $2$ amplitudes are as for the $K^+$ decay discussed in section~\ref{subsec:kto3pi}.

We use the same notation for the isospin-basis amplitudes as in eq.~(\ref{eq:Kamplitudes}), but now add
\begin{equation}
\mathbf T^{[0]}_{K3\pi}(p_1,p_2,p_3) \equiv i \cT_a^{[0]}(p_1,p_2,p_3)\,,
\end{equation}
where the subscript ``$a$'' denotes the antisymmetric irrep of $S_3$. Using $\cR$ in eq.~(\ref{eq:Rnumerical}) and the redundancy result eq.~(\ref{eq:redundancy}) we obtain the relation between isospin amplitudes and those for pions of definite charge,
\begin{align}
\cT^{000}(p_1,p_2,p_3) &= - \frac{3}{\sqrt{15}} \cT^{[1]}_s(p_1,p_2,p_3)
\\
\begin{split}
\cT^{+-0}(p_1,p_2,p_3) &=
\frac{1}{\sqrt{15}} \cT^{[1]}_s(p_1,p_2,p_3)
+ \frac{2}{3} \cT^{[2]}_{d,1}(p_1,p_2,p_3)
+ \frac{1}{3} \cT^{[2]}_{d,1}(p_1,p_3,p_2)
\\&\quad
+ \frac1{\sqrt3} \cT^{[1]}_{d,1}(p_1,p_3,p_2)
+\frac1{\sqrt6} \cT^{[0]}_a(p_1,p_2,p_3) \,.
\end{split}
\end{align}

\bibliographystyle{JHEP}
\bibliography{ref.bib}

\providecommand{\href}[2]{#2}\begingroup\raggedright\begin{thebibliography}{10}

\bibitem{Briceno:2012rv}
R.~A. Brice\~no and Z.~Davoudi, \emph{{Three-particle scattering amplitudes
  from a finite volume formalism}},
  \href{https://doi.org/10.1103/PhysRevD.87.094507}{\emph{Phys. Rev.}
  {\bfseries D87} (2013) 094507}
  [\href{https://arxiv.org/abs/1212.3398}{{\ttfamily 1212.3398}}].

\bibitem{Polejaeva:2012ut}
K.~Polejaeva and A.~Rusetsky, \emph{{Three particles in a finite volume}},
  \href{https://doi.org/10.1140/epja/i2012-12067-8}{\emph{Eur.\ Phys.\ J.\ A}
  {\bfseries 48} (2012) 67} [\href{https://arxiv.org/abs/1203.1241}{{\ttfamily
  1203.1241}}].

\bibitem{Hansen:2014eka}
M.~T. Hansen and S.~R. Sharpe, \emph{{Relativistic, model-independent,
  three-particle quantization condition}},
  \href{https://doi.org/10.1103/PhysRevD.90.116003}{\emph{Phys. Rev.}
  {\bfseries D90} (2014) 116003}
  [\href{https://arxiv.org/abs/1408.5933}{{\ttfamily 1408.5933}}].

\bibitem{Hansen:2015zga}
M.~T. Hansen and S.~R. Sharpe, \emph{{Expressing the three-particle
  finite-volume spectrum in terms of the three-to-three scattering amplitude}},
  \href{https://doi.org/10.1103/PhysRevD.92.114509}{\emph{Phys. Rev.}
  {\bfseries D92} (2015) 114509}
  [\href{https://arxiv.org/abs/1504.04248}{{\ttfamily 1504.04248}}].

\bibitem{Briceno:2017tce}
R.~A. Brice\~no, M.~T. Hansen and S.~R. Sharpe, \emph{{Relating the
  finite-volume spectrum and the two-and-three-particle $S$ matrix for
  relativistic systems of identical scalar particles}},
  \href{https://doi.org/10.1103/PhysRevD.95.074510}{\emph{Phys. Rev.}
  {\bfseries D95} (2017) 074510}
  [\href{https://arxiv.org/abs/1701.07465}{{\ttfamily 1701.07465}}].

\bibitem{Hammer:2017uqm}
H.-W. Hammer, J.-Y. Pang and A.~Rusetsky, \emph{{Three-particle quantization
  condition in a finite volume: 1. The role of the three-particle force}},
  \href{https://doi.org/10.1007/JHEP09(2017)109}{\emph{JHEP} {\bfseries 09}
  (2017) 109} [\href{https://arxiv.org/abs/1706.07700}{{\ttfamily
  1706.07700}}].

\bibitem{Hammer:2017kms}
H.~W. Hammer, J.~Y. Pang and A.~Rusetsky, \emph{{Three particle quantization
  condition in a finite volume: 2. general formalism and the analysis of
  data}}, \href{https://doi.org/10.1007/JHEP10(2017)115}{\emph{JHEP} {\bfseries
  10} (2017) 115} [\href{https://arxiv.org/abs/1707.02176}{{\ttfamily
  1707.02176}}].

\bibitem{Mai:2017bge}
M.~Mai and M.~{D\"oring}, \emph{{Three-body Unitarity in the Finite Volume}},
  \href{https://doi.org/10.1140/epja/i2017-12440-1}{\emph{Eur. Phys. J.}
  {\bfseries A53} (2017) 240}
  [\href{https://arxiv.org/abs/1709.08222}{{\ttfamily 1709.08222}}].

\bibitem{Briceno:2018aml}
R.~A. Brice\~no, M.~T. Hansen and S.~R. Sharpe, \emph{{Three-particle systems
  with resonant subprocesses in a finite volume}},
  \href{https://doi.org/10.1103/PhysRevD.99.014516}{\emph{Phys. Rev.}
  {\bfseries D99} (2019) 014516}
  [\href{https://arxiv.org/abs/1810.01429}{{\ttfamily 1810.01429}}].

\bibitem{Briceno:2018mlh}
R.~A. Brice\~no, M.~T. Hansen and S.~R. Sharpe, \emph{{Numerical study of the
  relativistic three-body quantization condition in the isotropic
  approximation}},
  \href{https://doi.org/10.1103/PhysRevD.98.014506}{\emph{Phys. Rev.}
  {\bfseries D98} (2018) 014506}
  [\href{https://arxiv.org/abs/1803.04169}{{\ttfamily 1803.04169}}].

\bibitem{Jackura:2019bmu}
A.~W. Jackura, S.~M. Dawid, C.~Fern\'andez-Ram\'irez, V.~Mathieu,
  M.~Mikhasenko, A.~Pilloni et~al., \emph{{Equivalence of three-particle
  scattering formalisms}},
  \href{https://doi.org/10.1103/PhysRevD.100.034508}{\emph{Phys. Rev.}
  {\bfseries D100} (2019) 034508}
  [\href{https://arxiv.org/abs/1905.12007}{{\ttfamily 1905.12007}}].

\bibitem{Blanton:2019igq}
T.~D. Blanton, F.~Romero-L\'opez and S.~R. Sharpe, \emph{{Implementing the
  three-particle quantization condition including higher partial waves}},
  \href{https://doi.org/10.1007/JHEP03(2019)106}{\emph{JHEP} {\bfseries 03}
  (2019) 106} [\href{https://arxiv.org/abs/1901.07095}{{\ttfamily
  1901.07095}}].

\bibitem{Briceno:2019muc}
R.~A. Briceño, M.~T. Hansen, S.~R. Sharpe and A.~P. Szczepaniak,
  \emph{{Unitarity of the infinite-volume three-particle scattering amplitude
  arising from a finite-volume formalism}},
  \href{https://doi.org/10.1103/PhysRevD.100.054508}{\emph{Phys. Rev.}
  {\bfseries D100} (2019) 054508}
  [\href{https://arxiv.org/abs/1905.11188}{{\ttfamily 1905.11188}}].

\bibitem{Hansen:2019nir}
M.~T. Hansen and S.~R. Sharpe, \emph{{Lattice QCD and Three-particle Decays of
  Resonances}},
  \href{https://doi.org/10.1146/annurev-nucl-101918-023723}{\emph{Ann. Rev.
  Nucl. Part. Sci.} {\bfseries 69} (2019) 65}
  [\href{https://arxiv.org/abs/1901.00483}{{\ttfamily 1901.00483}}].

\bibitem{Romero-Lopez:2019qrt}
F.~Romero-López, S.~R. Sharpe, T.~D. Blanton, R.~A. Briceño and M.~T. Hansen,
  \emph{{Numerical exploration of three relativistic particles in a finite
  volume including two-particle resonances and bound states}},
  \href{https://doi.org/10.1007/JHEP10(2019)007}{\emph{JHEP} {\bfseries 10}
  (2019) 007} [\href{https://arxiv.org/abs/1908.02411}{{\ttfamily
  1908.02411}}].

\bibitem{Blanton:2020gha}
T.~D. Blanton and S.~R. Sharpe, \emph{{Alternative derivation of the
  relativistic three-particle quantization condition}},
  \href{https://doi.org/10.1103/PhysRevD.102.054520}{\emph{Phys. Rev. D}
  {\bfseries 102} (2020) 054520}
  [\href{https://arxiv.org/abs/2007.16188}{{\ttfamily 2007.16188}}].

\bibitem{Blanton:2020jnm}
T.~D. Blanton and S.~R. Sharpe, \emph{{Equivalence of relativistic
  three-particle quantization conditions}},
  \href{https://doi.org/10.1103/PhysRevD.102.054515}{\emph{Phys. Rev. D}
  {\bfseries 102} (2020) 054515}
  [\href{https://arxiv.org/abs/2007.16190}{{\ttfamily 2007.16190}}].

\bibitem{Hansen:2020zhy}
M.~T. Hansen, F.~Romero-L\'opez and S.~R. Sharpe, \emph{{Generalizing the
  relativistic quantization condition to include all three-pion isospin
  channels}}, \href{https://doi.org/10.1007/JHEP07(2020)047}{\emph{JHEP}
  {\bfseries 07} (2020) 047}
  [\href{https://arxiv.org/abs/2003.10974}{{\ttfamily 2003.10974}}].

\bibitem{Blanton:2020gmf}
T.~D. Blanton and S.~R. Sharpe, \emph{{Relativistic three-particle quantization
  condition for nondegenerate scalars}},
  \href{https://arxiv.org/abs/2011.05520}{{\ttfamily 2011.05520}}.

\bibitem{Muller:2020vtt}
F.~M\"uller, A.~Rusetsky and T.~Yu, \emph{{Finite-volume energy shift of the
  three-pion ground state}},
  \href{https://arxiv.org/abs/2011.14178}{{\ttfamily 2011.14178}}.

\bibitem{Mai:2018djl}
M.~Mai and M.~D{\"o}ring, \emph{{Finite-Volume Spectrum of $\pi^+\pi^+$ and
  $\pi^+\pi^+\pi^+$ Systems}},
  \href{https://doi.org/10.1103/PhysRevLett.122.062503}{\emph{Phys. Rev. Lett.}
  {\bfseries 122} (2019) 062503}
  [\href{https://arxiv.org/abs/1807.04746}{{\ttfamily 1807.04746}}].

\bibitem{Horz:2019rrn}
B.~Hörz and A.~Hanlon, \emph{{Two- and three-pion finite-volume spectra at
  maximal isospin from lattice QCD}},
  \href{https://doi.org/10.1103/PhysRevLett.123.142002}{\emph{Phys. Rev. Lett.}
  {\bfseries 123} (2019) 142002}
  [\href{https://arxiv.org/abs/1905.04277}{{\ttfamily 1905.04277}}].

\bibitem{Blanton:2019vdk}
T.~D. Blanton, F.~Romero-L\'opez and S.~R. Sharpe, \emph{{$I = 3$ three-pion
  scattering amplitude from lattice QCD}},
  \href{https://doi.org/10.1103/PhysRevLett.124.032001}{\emph{Phys. Rev. Lett.}
  {\bfseries 124} (2020) 032001}
  [\href{https://arxiv.org/abs/1909.02973}{{\ttfamily 1909.02973}}].

\bibitem{Culver:2019vvu}
C.~Culver, M.~Mai, R.~Brett, A.~Alexandru and M.~D\"oring, \emph{{Three body
  spectrum from lattice QCD}},
  \href{https://doi.org/10.1103/PhysRevD.101.114507}{\emph{Phys. Rev. D}
  {\bfseries 101} (2020) 114507}
  [\href{https://arxiv.org/abs/1911.09047}{{\ttfamily 1911.09047}}].

\bibitem{Mai:2019fba}
M.~Mai, M.~Döring, C.~Culver and A.~Alexandru, \emph{{Three-body unitarity
  versus finite-volume $\pi^+\pi^+\pi^+$ spectrum from lattice QCD}},
  \href{https://doi.org/10.1103/PhysRevD.101.054510}{\emph{Phys.\ Rev.\ D}
  {\bfseries 101} (2020) 054510}
  [\href{https://arxiv.org/abs/1909.05749}{{\ttfamily 1909.05749}}].

\bibitem{Fischer:2020jzp}
M.~Fischer, B.~Kostrzewa, L.~Liu, F.~Romero-L\'opez, M.~Ueding and C.~Urbach,
  \emph{{Scattering of two and three physical pions at maximal isospin from
  lattice QCD}},  \href{https://arxiv.org/abs/2008.03035}{{\ttfamily
  2008.03035}}.

\bibitem{Hansen:2020otl}
M.~T. Hansen, R.~A. Brice\~no, R.~G. Edwards, C.~E. Thomas and D.~J. Wilson,
  \emph{{The energy-dependent $\pi^+ \pi^+ \pi^+$ scattering amplitude from
  QCD}}, \href{https://doi.org/10.1103/PhysRevLett.126.012001}{\emph{Phys. Rev.
  Lett.} {\bfseries 126} (2021) 012001}
  [\href{https://arxiv.org/abs/2009.04931}{{\ttfamily 2009.04931}}].

\bibitem{Alexandru:2020xqf}
A.~Alexandru, R.~Brett, C.~Culver, M.~D\"oring, D.~Guo, F.~X. Lee et~al.,
  \emph{{Finite-volume energy spectrum of the $K^-K^-K^-$ system}},
  \href{https://doi.org/10.1103/PhysRevD.102.114523}{\emph{Phys. Rev. D}
  {\bfseries 102} (2021) 114523}
  [\href{https://arxiv.org/abs/2009.12358}{{\ttfamily 2009.12358}}].

\bibitem{Brett:2021wyd}
R.~Brett, C.~Culver, M.~Mai, A.~Alexandru, M.~D\"oring and F.~X. Lee,
  \emph{{Three-body interactions from the finite-volume QCD spectrum}},
  \href{https://arxiv.org/abs/2101.06144}{{\ttfamily 2101.06144}}.

\bibitem{Romero-Lopez:2018rcb}
F.~Romero-López, A.~Rusetsky and C.~Urbach, \emph{{Two- and three-body
  interactions in $\varphi ^4$ theory from lattice simulations}},
  \href{https://doi.org/10.1140/epjc/s10052-018-6325-8}{\emph{Eur. Phys. J.}
  {\bfseries C78} (2018) 846}
  [\href{https://arxiv.org/abs/1806.02367}{{\ttfamily 1806.02367}}].

\bibitem{Romero-Lopez:2020rdq}
F.~Romero-L\'opez, A.~Rusetsky, N.~Schlage and C.~Urbach, \emph{{Relativistic
  $N$-particle energy shift in finite volume}},
  \href{https://arxiv.org/abs/2010.11715}{{\ttfamily 2010.11715}}.

\bibitem{Lellouch:2000pv}
L.~Lellouch and M.~Luscher, \emph{{Weak transition matrix elements from finite
  volume correlation functions}},
  \href{https://doi.org/10.1007/s002200100410}{\emph{Commun. Math. Phys.}
  {\bfseries 219} (2001) 31}
  [\href{https://arxiv.org/abs/hep-lat/0003023}{{\ttfamily hep-lat/0003023}}].

\bibitem{Lin:2001ek}
C.~Lin, G.~Martinelli, C.~T. Sachrajda and M.~Testa, \emph{{$K\to\pi\pi$ decays
  in a finite volume}},
  \href{https://doi.org/10.1016/S0550-3213(01)00495-3}{\emph{Nucl. Phys. B}
  {\bfseries 619} (2001) 467}
  [\href{https://arxiv.org/abs/hep-lat/0104006}{{\ttfamily hep-lat/0104006}}].

\bibitem{Detmold:2004qn}
W.~Detmold and M.~J. Savage, \emph{{Electroweak matrix elements in the two
  nucleon sector from lattice QCD}},
  \href{https://doi.org/10.1016/j.nuclphysa.2004.07.007}{\emph{Nucl. Phys. A}
  {\bfseries 743} (2004) 170}
  [\href{https://arxiv.org/abs/hep-lat/0403005}{{\ttfamily hep-lat/0403005}}].

\bibitem{Kim:2005gf}
C.~h. Kim, C.~T. Sachrajda and S.~R. Sharpe, \emph{{Finite-volume effects for
  two-hadron states in moving frames}},
  \href{https://doi.org/10.1016/j.nuclphysb.2005.08.029}{\emph{Nucl. Phys.}
  {\bfseries B727} (2005) 218}
  [\href{https://arxiv.org/abs/hep-lat/0507006}{{\ttfamily hep-lat/0507006}}].

\bibitem{Christ:2005gi}
N.~H. Christ, C.~Kim and T.~Yamazaki, \emph{{Finite volume corrections to the
  two-particle decay of states with non-zero momentum}},
  \href{https://doi.org/10.1103/PhysRevD.72.114506}{\emph{Phys. Rev. D}
  {\bfseries 72} (2005) 114506}
  [\href{https://arxiv.org/abs/hep-lat/0507009}{{\ttfamily hep-lat/0507009}}].

\bibitem{Meyer:2011um}
H.~B. Meyer, \emph{{Lattice QCD and the Timelike Pion Form Factor}},
  \href{https://doi.org/10.1103/PhysRevLett.107.072002}{\emph{Phys. Rev. Lett.}
  {\bfseries 107} (2011) 072002}
  [\href{https://arxiv.org/abs/1105.1892}{{\ttfamily 1105.1892}}].

\bibitem{Hansen:2012tf}
M.~T. Hansen and S.~R. Sharpe, \emph{{Multiple-channel generalization of
  Lellouch-Lüscher formula}},
  \href{https://doi.org/10.1103/PhysRevD.86.016007}{\emph{Phys.Rev.} {\bfseries
  D86} (2012) 016007} [\href{https://arxiv.org/abs/1204.0826}{{\ttfamily
  1204.0826}}].

\bibitem{Briceno:2012yi}
R.~A. Brice{\~n}o and Z.~Davoudi, \emph{{Moving multichannel systems in a
  finite volume with application to proton-proton fusion}},
  \href{https://doi.org/10.1103/PhysRevD.88.094507}{\emph{Phys. Rev.}
  {\bfseries D88} (2013) 094507}
  [\href{https://arxiv.org/abs/1204.1110}{{\ttfamily 1204.1110}}].

\bibitem{Bernard:2012bi}
V.~Bernard, D.~Hoja, U.~Mei{\ss}ner and A.~Rusetsky, \emph{{Matrix elements of
  unstable states}}, \href{https://doi.org/10.1007/JHEP09(2012)023}{\emph{JHEP}
  {\bfseries 09} (2012) 023} [\href{https://arxiv.org/abs/1205.4642}{{\ttfamily
  1205.4642}}].

\bibitem{Agadjanov:2014kha}
A.~Agadjanov, V.~Bernard, U.~Mei{\ss}ner and A.~Rusetsky, \emph{{A framework
  for the calculation of the $\Delta N \gamma^*$ transition form factors on the
  lattice}}, \href{https://doi.org/10.1016/j.nuclphysb.2014.07.023}{\emph{Nucl.
  Phys. B} {\bfseries 886} (2014) 1199}
  [\href{https://arxiv.org/abs/1405.3476}{{\ttfamily 1405.3476}}].

\bibitem{Briceno:2014uqa}
R.~A. Brice\~no, M.~T. Hansen and A.~Walker-Loud, \emph{{Multichannel 1
  $\rightarrow$ 2 transition amplitudes in a finite volume}},
  \href{https://doi.org/10.1103/PhysRevD.91.034501}{\emph{Phys. Rev.}
  {\bfseries D91} (2015) 034501}
  [\href{https://arxiv.org/abs/1406.5965}{{\ttfamily 1406.5965}}].

\bibitem{Feng:2014gba}
X.~Feng, S.~Aoki, S.~Hashimoto and T.~Kaneko, \emph{{Timelike pion form factor
  in lattice QCD}},
  \href{https://doi.org/10.1103/PhysRevD.91.054504}{\emph{Phys. Rev.}
  {\bfseries D91} (2015) 054504}
  [\href{https://arxiv.org/abs/1412.6319}{{\ttfamily 1412.6319}}].

\bibitem{Briceno:2015csa}
R.~A. Brice\~no and M.~T. Hansen, \emph{{Multichannel 0 $\to$ 2 and 1 $\to$ 2
  transition amplitudes for arbitrary spin particles in a finite volume}},
  \href{https://doi.org/10.1103/PhysRevD.92.074509}{\emph{Phys. Rev.}
  {\bfseries D92} (2015) 074509}
  [\href{https://arxiv.org/abs/1502.04314}{{\ttfamily 1502.04314}}].

\bibitem{Briceno:2015tza}
R.~A. Brice{\~n}o and M.~T. Hansen, \emph{{Relativistic, model-independent,
  multichannel $2\to 2$ transition amplitudes in a finite volume}},
  \href{https://doi.org/10.1103/PhysRevD.94.013008}{\emph{Phys. Rev. D}
  {\bfseries 94} (2016) 013008}
  [\href{https://arxiv.org/abs/1509.08507}{{\ttfamily 1509.08507}}].

\bibitem{Baroni:2018iau}
A.~Baroni, R.~A. Brice{\~n}o, M.~T. Hansen and F.~G. Ortega-Gama, \emph{{Form
  factors of two-hadron states from a covariant finite-volume formalism}},
  \href{https://doi.org/10.1103/PhysRevD.100.034511}{\emph{Phys. Rev. D}
  {\bfseries 100} (2019) 034511}
  [\href{https://arxiv.org/abs/1812.10504}{{\ttfamily 1812.10504}}].

\bibitem{Briceno:2019nns}
R.~A. Brice{\~n}o, M.~T. Hansen and A.~W. Jackura, \emph{{Consistency checks
  for two-body finite-volume matrix elements: I. Conserved currents and bound
  states}}, \href{https://doi.org/10.1103/PhysRevD.100.114505}{\emph{Phys. Rev.
  D} {\bfseries 100} (2019) 114505}
  [\href{https://arxiv.org/abs/1909.10357}{{\ttfamily 1909.10357}}].

\bibitem{Briceno:2020xxs}
R.~A. Brice{\~n}o, M.~T. Hansen and A.~W. Jackura, \emph{{Consistency checks
  for two-body finite-volume matrix elements: II. Perturbative systems}},
  \href{https://doi.org/10.1103/PhysRevD.101.094508}{\emph{Phys. Rev. D}
  {\bfseries 101} (2020) 094508}
  [\href{https://arxiv.org/abs/2002.00023}{{\ttfamily 2002.00023}}].

\bibitem{Feng:2020nqj}
X.~Feng, L.-C. Jin, Z.-Y. Wang and Z.~Zhang, \emph{{Finite-volume formalism in
  the $2 \xrightarrow[]{H_I+H_I} 2$ transition: an application to the lattice
  QCD calculation of double beta decays}},
  \href{https://arxiv.org/abs/2005.01956}{{\ttfamily 2005.01956}}.

\bibitem{Bernecker:2011gh}
D.~Bernecker and H.~B. Meyer, \emph{{Vector Correlators in Lattice QCD: Methods
  and applications}},
  \href{https://doi.org/10.1140/epja/i2011-11148-6}{\emph{Eur. Phys. J. A}
  {\bfseries 47} (2011) 148} [\href{https://arxiv.org/abs/1107.4388}{{\ttfamily
  1107.4388}}].

\bibitem{Meyer:2018til}
H.~B. Meyer and H.~Wittig, \emph{{Lattice QCD and the anomalous magnetic moment
  of the muon}}, \href{https://doi.org/10.1016/j.ppnp.2018.09.001}{\emph{Prog.
  Part. Nucl. Phys.} {\bfseries 104} (2019) 46}
  [\href{https://arxiv.org/abs/1807.09370}{{\ttfamily 1807.09370}}].

\bibitem{Muller:2020wjo}
F.~M\"uller and A.~Rusetsky, \emph{{On the three-particle analog of the
  Lellouch-L\"uscher formula}},
  \href{https://arxiv.org/abs/2012.13957}{{\ttfamily 2012.13957}}.

\bibitem{Khuri:1960zz}
N.~Khuri and S.~Treiman, \emph{{Pion-Pion Scattering and $K^{\pm}\to 3\pi$
  Decay}}, \href{https://doi.org/10.1103/PhysRev.119.1115}{\emph{Phys. Rev.}
  {\bfseries 119} (1960) 1115}.

\bibitem{Jackura:2020bsk}
A.~W. Jackura, R.~A. Brice\~no, S.~M. Dawid, M.~H.~E. Islam and C.~McCarty,
  \emph{{Solving relativistic three-body integral equations in the presence of
  bound states}},  \href{https://arxiv.org/abs/2010.09820}{{\ttfamily
  2010.09820}}.

\bibitem{Hoferichter:2014vra}
M.~Hoferichter, B.~Kubis, S.~Leupold, F.~Niecknig and S.~P. Schneider,
  \emph{{Dispersive analysis of the pion transition form factor}},
  \href{https://doi.org/10.1140/epjc/s10052-014-3180-0}{\emph{Eur. Phys. J. C}
  {\bfseries 74} (2014) 3180}
  [\href{https://arxiv.org/abs/1410.4691}{{\ttfamily 1410.4691}}].

\bibitem{Hoferichter:2018dmo}
M.~Hoferichter, B.-L. Hoid, B.~Kubis, S.~Leupold and S.~P. Schneider,
  \emph{{Pion-pole contribution to hadronic light-by-light scattering in the
  anomalous magnetic moment of the muon}},
  \href{https://doi.org/10.1103/PhysRevLett.121.112002}{\emph{Phys. Rev. Lett.}
  {\bfseries 121} (2018) 112002}
  [\href{https://arxiv.org/abs/1805.01471}{{\ttfamily 1805.01471}}].

\bibitem{Hoferichter:2018kwz}
M.~Hoferichter, B.-L. Hoid, B.~Kubis, S.~Leupold and S.~P. Schneider,
  \emph{{Dispersion relation for hadronic light-by-light scattering: pion
  pole}}, \href{https://doi.org/10.1007/JHEP10(2018)141}{\emph{JHEP} {\bfseries
  10} (2018) 141} [\href{https://arxiv.org/abs/1808.04823}{{\ttfamily
  1808.04823}}].

\bibitem{Hoferichter:2019mqg}
M.~Hoferichter, B.-L. Hoid and B.~Kubis, \emph{{Three-pion contribution to
  hadronic vacuum polarization}},
  \href{https://doi.org/10.1007/JHEP08(2019)137}{\emph{JHEP} {\bfseries 08}
  (2019) 137} [\href{https://arxiv.org/abs/1907.01556}{{\ttfamily
  1907.01556}}].

\bibitem{Hoid:2020xjs}
B.-L. Hoid, M.~Hoferichter and B.~Kubis, \emph{{Hadronic vacuum polarization
  and vector-meson resonance parameters from $e^+e^-\rightarrow \pi
  ^0\gamma$}},
  \href{https://doi.org/10.1140/epjc/s10052-020-08550-2}{\emph{Eur. Phys. J. C}
  {\bfseries 80} (2020) 988}
  [\href{https://arxiv.org/abs/2007.12696}{{\ttfamily 2007.12696}}].

\bibitem{Zyla:2020zbs}
{\scshape Particle Data Group} collaboration, P.~Zyla et~al., \emph{{Review of
  Particle Physics}}, \href{https://doi.org/10.1093/ptep/ptaa104}{\emph{PTEP}
  {\bfseries 2020} (2020) 083C01}.

\bibitem{Gan:2020aco}
L.~Gan, B.~Kubis, E.~Passemar and S.~Tulin, \emph{{Precision tests of
  fundamental physics with $\eta$ and $\eta^\prime$ mesons}},
  \href{https://arxiv.org/abs/2007.00664}{{\ttfamily 2007.00664}}.

\bibitem{deDivitiis:2011eh}
G.~de~Divitiis et~al., \emph{{Isospin breaking effects due to the up-down mass
  difference in Lattice QCD}},
  \href{https://doi.org/10.1007/JHEP04(2012)124}{\emph{JHEP} {\bfseries 04}
  (2012) 124} [\href{https://arxiv.org/abs/1110.6294}{{\ttfamily 1110.6294}}].

\bibitem{deDivitiis:2013xla}
{\scshape RM123} collaboration, G.~de~Divitiis, R.~Frezzotti, V.~Lubicz,
  G.~Martinelli, R.~Petronzio, G.~Rossi et~al., \emph{{Leading isospin breaking
  effects on the lattice}},
  \href{https://doi.org/10.1103/PhysRevD.87.114505}{\emph{Phys. Rev. D}
  {\bfseries 87} (2013) 114505}
  [\href{https://arxiv.org/abs/1303.4896}{{\ttfamily 1303.4896}}].

\bibitem{Colangelo:2018jxw}
G.~Colangelo, S.~Lanz, H.~Leutwyler and E.~Passemar, \emph{{Dispersive analysis
  of $\eta \rightarrow 3 \pi $}},
  \href{https://doi.org/10.1140/epjc/s10052-018-6377-9}{\emph{Eur. Phys. J. C}
  {\bfseries 78} (2018) 947}
  [\href{https://arxiv.org/abs/1807.11937}{{\ttfamily 1807.11937}}].

\bibitem{Kampf:2019bkf}
K.~Kampf, M.~Knecht, J.~Novotn\'y and M.~Zdr\'ahal, \emph{{Dispersive
  construction of two-loop $P \to \pi\pi\pi$ $(P=K,\eta)$ amplitudes}},
  \href{https://doi.org/10.1103/PhysRevD.101.074043}{\emph{Phys. Rev. D}
  {\bfseries 101} (2020) 074043}
  [\href{https://arxiv.org/abs/1911.11762}{{\ttfamily 1911.11762}}].

\bibitem{Bai:2015nea}
{\scshape RBC, UKQCD} collaboration, Z.~Bai et~al., \emph{{Standard Model
  Prediction for Direct CP Violation in $K \to \pi\pi$ Decay}},
  \href{https://doi.org/10.1103/PhysRevLett.115.212001}{\emph{Phys. Rev. Lett.}
  {\bfseries 115} (2015) 212001}
  [\href{https://arxiv.org/abs/1505.07863}{{\ttfamily 1505.07863}}].

\bibitem{Blum:2015ywa}
T.~Blum et~al., \emph{{$K \rightarrow \pi\pi$ $\Delta I=3/2$ decay amplitude in
  the continuum limit}},
  \href{https://doi.org/10.1103/PhysRevD.91.074502}{\emph{Phys. Rev. D}
  {\bfseries 91} (2015) 074502}
  [\href{https://arxiv.org/abs/1502.00263}{{\ttfamily 1502.00263}}].

\bibitem{Abbott:2020hxn}
{\scshape RBC, UKQCD} collaboration, R.~Abbott et~al., \emph{{Direct CP
  violation and the $\Delta I=1/2$ rule in $K\to\pi\pi$ decay from the standard
  model}}, \href{https://doi.org/10.1103/PhysRevD.102.054509}{\emph{Phys. Rev.
  D} {\bfseries 102} (2020) 054509}
  [\href{https://arxiv.org/abs/2004.09440}{{\ttfamily 2004.09440}}].

\bibitem{Cirigliano:2011ny}
V.~Cirigliano, G.~Ecker, H.~Neufeld, A.~Pich and J.~Portoles, \emph{{Kaon
  Decays in the Standard Model}},
  \href{https://doi.org/10.1103/RevModPhys.84.399}{\emph{Rev. Mod. Phys.}
  {\bfseries 84} (2012) 399} [\href{https://arxiv.org/abs/1107.6001}{{\ttfamily
  1107.6001}}].

\bibitem{Batley:2007aa}
{\scshape NA48/2} collaboration, J.~Batley et~al., \emph{{Search for direct CP
  violating charge asymmetries in K+- ---\ensuremath{>} pi+- pi+ pi- and K+-
  ---\ensuremath{>} pi+- pi0 pi0 decays}},
  \href{https://doi.org/10.1140/epjc/s10052-007-0456-7}{\emph{Eur. Phys. J. C}
  {\bfseries 52} (2007) 875} [\href{https://arxiv.org/abs/0707.0697}{{\ttfamily
  0707.0697}}].

\bibitem{Batley:2010fj}
{\scshape NA48/2} collaboration, J.~Batley et~al., \emph{{Empirical
  parameterization of the K+- -\ensuremath{>} pi+- pi0 pi0 decay Dalitz plot}},
  \href{https://doi.org/10.1016/j.physletb.2010.02.036}{\emph{Phys. Lett. B}
  {\bfseries 686} (2010) 101}
  [\href{https://arxiv.org/abs/1004.1005}{{\ttfamily 1004.1005}}].

\bibitem{Gamiz:2003pi}
E.~Gamiz, J.~Prades and I.~Scimemi, \emph{{Charged kaon K ---\ensuremath{>} 3pi
  CP violating asymmetries at NLO in CHPT}},
  \href{https://doi.org/10.1088/1126-6708/2003/10/042}{\emph{JHEP} {\bfseries
  10} (2003) 042} [\href{https://arxiv.org/abs/hep-ph/0309172}{{\ttfamily
  hep-ph/0309172}}].

\bibitem{Prades:2007ud}
J.~Prades, \emph{{ChPT Progress on Non-Leptonic and Radiative Kaon Decays}},
  \href{https://doi.org/10.22323/1.046.0022}{\emph{PoS} {\bfseries KAON} (2008)
  022} [\href{https://arxiv.org/abs/0707.1789}{{\ttfamily 0707.1789}}].

\bibitem{Buchalla:1995vs}
G.~Buchalla, A.~J. Buras and M.~E. Lautenbacher, \emph{{Weak decays beyond
  leading logarithms}},
  \href{https://doi.org/10.1103/RevModPhys.68.1125}{\emph{Rev. Mod. Phys.}
  {\bfseries 68} (1996) 1125}
  [\href{https://arxiv.org/abs/hep-ph/9512380}{{\ttfamily hep-ph/9512380}}].

\bibitem{Blum:2001xb}
{\scshape RBC} collaboration, T.~Blum et~al., \emph{{Kaon matrix elements and
  CP violation from quenched lattice QCD: 1. The three flavor case}},
  \href{https://doi.org/10.1103/PhysRevD.68.114506}{\emph{Phys. Rev. D}
  {\bfseries 68} (2003) 114506}
  [\href{https://arxiv.org/abs/hep-lat/0110075}{{\ttfamily hep-lat/0110075}}].

\end{thebibliography}\endgroup

\end{document}